\documentclass[reprint,amsmath,amssymb,aip]{revtex4-2}
\usepackage{graphicx}
\usepackage{dcolumn}
\usepackage{bm}
\usepackage[utf8]{inputenc}
\usepackage[T1]{fontenc}
\usepackage{mathptmx}
\usepackage{etoolbox}

\makeatletter
\def\@email#1#2{
 \endgroup
 \patchcmd{\titleblock@produce}
  {\frontmatter@RRAPformat}
  {\frontmatter@RRAPformat{\produce@RRAP{*#1\href{mailto:#2}{#2}}}\frontmatter@RRAPformat}
  {}{}
}
\newcounter{suppfig}
\renewcommand{\thesuppfig}{S} 

\newcommand{\suppnumbering}{%
  \setcounter{figure}{0} 
  \renewcommand{\thefigure}{\thesuppfig\arabic{figure}} 
}

\newcommand{\regularnumbering}{%
  \renewcommand{\thefigure}{\arabic{figure}} 
}
\newcommand{\largeSection}[1]{%
    \begingroup%
    \LARGE%
    \section*{#1}%
    \endgroup%
}
\makeatother

\newcommand{\Bext}{$B_{\mathrm{ext}}$}

\begin{document}

\preprint{AIP/123-QED}

\title{Spin-Wave Self-Imaging: Experimental and Numerical Demonstration\\ of Caustic and Talbot-like Diffraction Patterns}

\author{Uladzislau Makartsou}
\email{ulamak@amu.edu.pl}
\affiliation{Institute of Spintronics and Quantum Information, Faculty of Physics, Adam Mickiewicz University, Uniwersytetu Poznańskiego 2, 61-614 Poznań, Poland}
 
\author{Mateusz Gołębiewski}
\affiliation{Institute of Spintronics and Quantum Information, Faculty of Physics, Adam Mickiewicz University, Uniwersytetu Poznańskiego 2, 61-614 Poznań, Poland}

\author{Urszula Guzowska}
\affiliation{Department of Physics of Magnetism, Faculty of Physics, University of Białystok, Konstantego Ciołkowskiego 1L, 15-245 Białystok, Poland}

\author{Alexander Stognij}
\affiliation{Scientific-Practical Materials Research Center at National Academy of Sciences of Belarus, Pietrusia Broŭki 19, 220072 Minsk, Belarus}

\author{Ryszard Gieniusz}
\affiliation{Department of Physics of Magnetism, Faculty of Physics, University of Białystok, Konstantego Ciołkowskiego 1L, 15-245 Białystok, Poland}

\author{Maciej Krawczyk}
\affiliation{Institute of Spintronics and Quantum Information, Faculty of Physics, Adam Mickiewicz University, Uniwersytetu Poznańskiego 2, 61-614 Poznań, Poland}

\date{\today}

\begin{abstract}
Extending the scope of the self-imaging phenomenon, traditionally associated with linear optics, to the domain of magnonics, this study presents the experimental demonstration and numerical analysis of spin-wave (SW) self-imaging in an in-plane magnetized yttrium iron garnet film. We explore this phenomenon using a setup in which a plane SW passes through a diffraction grating, and the resulting interference pattern is detected using Brillouin light scattering. We have varied the frequencies of the source dynamic magnetic field to discern the influence of the anisotropic dispersion relation and the caustic effect on the analyzed phenomenon. We found that at low frequencies and diffraction fields, the caustics determine the interference pattern. However, at large distances from the grating, when the waves of high diffraction order and number of slits contribute to the interference pattern, the self-imaging phenomenon and Talbot-like patterns are formed. This methodological approach not only sheds light on the behavior of SW interference under different conditions but also enhances our understanding of the SW self-imaging process in both isotropic and anisotropic media.
\end{abstract}

\maketitle

Spin waves (SWs) represent coherent magnetization disturbances that propagate in magnetic materials as waveforms. In ferromagnetic materials, SW dynamics are shaped by a blend of strong isotropic exchange and anisotropic magnetostatic interactions~\cite{Stancil2009,gurevich1996book}. Particularly in thin films, the magnetostatic interactions render SW properties highly sensitive to the orientation of magnetization relative to the film plane, as well as to the alignment of the propagation direction with the static magnetization vector~\cite{Garcia-Sanchez2015NarrowWalls, Wagner2016MagneticNanochannels, Duerr2012EnhancedWaveguide, Lan2015Spin-WaveDiode}. This interplay of factors makes the study of SWs interesting, endowing them with distinct properties uncommon in other wave types, such as negative group velocity, the formation of caustics~\cite{Lock2018}, and dynamic reconfigurability control~\cite{Pirro2021AdvancesMagnonics}.

The governing equations for SW propagation diverge from those of electromagnetic and acoustic waves, thereby each analogy, such as those found in SW graded index lenses~\cite{graded_index_lenses,Kiechle2023}, SW Luneburg lenses~\cite{Luneberg}, and SW Fourier optics~\cite{SW_Fourier}, necessitates solving the Landau–Lifshitz equation. Notable and related advances also include the numerical or experimental demonstration of phenomena such as self-focusing of SWs~\cite{Dem08}, SW diffraction on gratings~\cite{Man12}, and the formation of SW beams~\cite{Kho04, gieniusz2017switching, korner2017excitation}. Ferromagnetic films with a line of nanodots, analogous to those in this paper, have also been used to observe the phenomenon known in literature as total non-reflection of SWs~\cite{Gieniusz2014}.

The self-imaging effect, often referred to as the Talbot effect, first observed in the 19th century for light~\cite{Talbot36}, and later elucidated in Ref.~[\onlinecite{Rayleigh81}], has recently experienced a renewed research interest, as outlined in Ref.~[\onlinecite{Wen13}] and associated references. When a plane wave passes through a system of periodically spaced obstacles, it interferes, creating a characteristic diffraction pattern, reproducing the obstacles image at specific distances from the input. 

Its applications have been diverse, ranging from enhancing x-ray imaging~\cite{Bravin_2012} to advancing lithographic patterning~\cite{Sat14,Zho16,Vet18}, and even extending to the realization of certain physical models and computing scenarios~\cite{Big08,Far15,Saw18}. Beyond electromagnetic waves, the Talbot effect has been demonstrated in diverse mediums including plasmons~\cite{Den07}, fluid waves~\cite{Sun18, Bakman19}, and exciton-polaritons~\cite{Gao16}. Theoretical explorations suggest that this effect is also feasible for SWs~\cite{Golebiewski2020,Golebiewski_multimode}, with proposed logic scenarios in the magnonic domain~\cite{Golebiewski_lookup}.

In this paper, we take a step forward by experimentally and numerically demonstrating the self-imaging effect resulting from the diffraction of SWs on a one-dimensional antidot array, with dimensions comparable to the SW wavelength, in a yttrium iron garnet (YIG) film in the Damon-Eshabach (DE) configuration. The simulations, performed at various frequencies, different antidot periods, and two antidot shapes, i.e., circular and square, and juxtaposed with experimental analogs, allow an understanding of the interference patterns in the transition of the caustic and self-imaging effects.

\begin{figure}[htp]
\centering
\includegraphics[width=\linewidth]{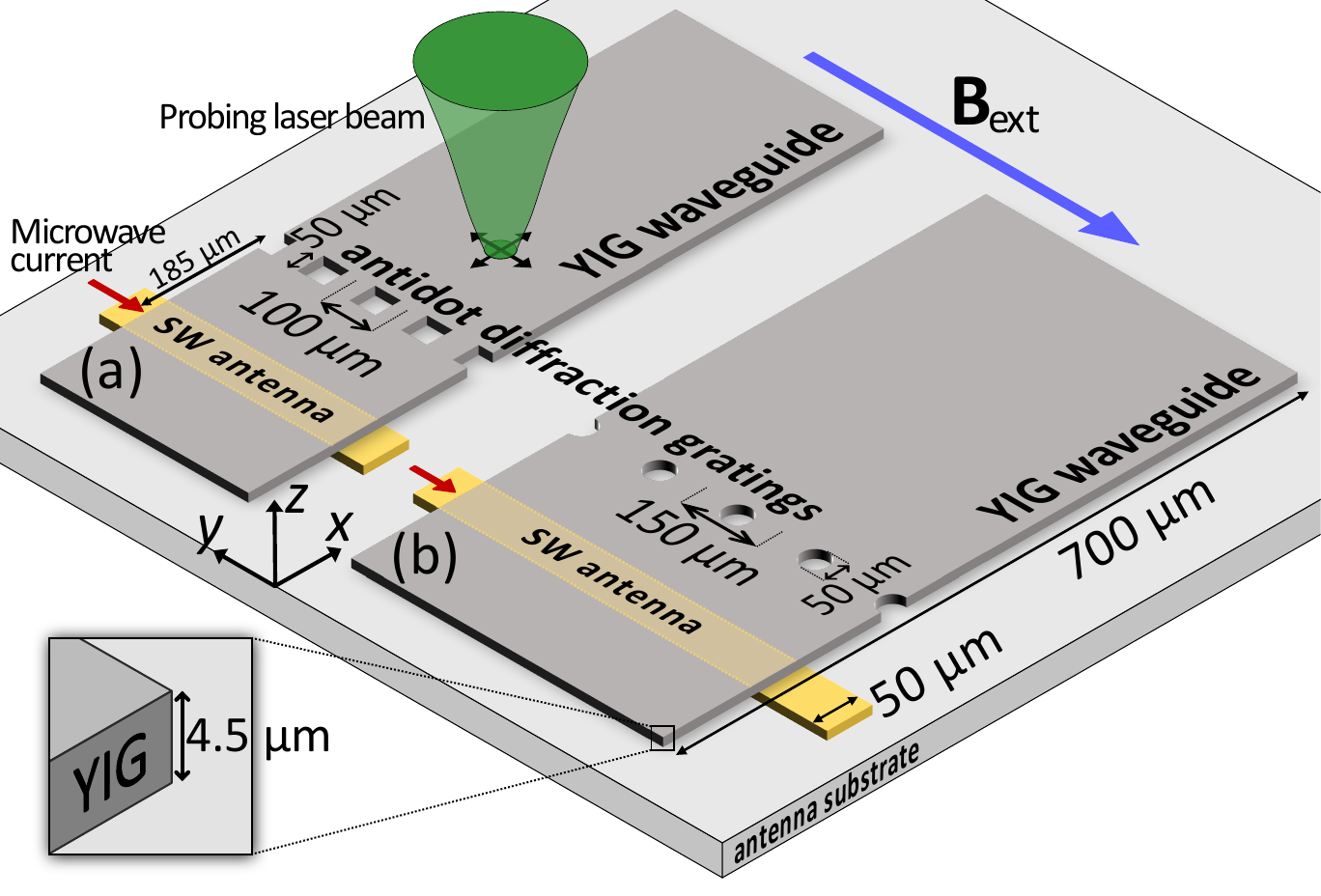}
\caption{\label{fig:geom}
Schematic representations of two YIG-samples and BLS measurement configuration used in the research, with the major dimensions marked. Image (a) shows the system with a diffraction grating made of square-shaped antidots, while (b) shows one made of circular antidots. In all experimental measurements and finite-width simulations, the square and circular antidot arrays contain 5 and 10 elements, respectively.}
\end{figure}

The samples measured were monocrystalline YIG, 4.5~\textmu m thick films (see Fig.~\ref{fig:geom}) grown by liquid phase epitaxy on transparent gadolinium gallium garnet substrates. A one-dimensional array of antidots was chemically etched on the surface of the films, functioning as a diffraction grating. In the two systems analyzed, we make use of an 5-element array of square antidots [$50 \times 50$~\textmu m -- scheme in Fig.~\ref{fig:geom}(a)] with period $d=100$~\textmu m, and an 10-element array of circular antidots [diameter equal to 50~\textmu m -- scheme in Fig.~\ref{fig:geom}(b)] with period $d=150$~\textmu m. The samples were magnetized by the external magnetic field ${\bf B}_{\text{ext}}$ directed along the line of the antidots ($y$-axis). We use 36~mT and 98~mT fields for the samples with square and circular antidots, respectively. Magnetostatic SWs were excited using a 50~\textmu m wide microstrip antenna deposited on a light-opaque dielectric substrate, below the YIG film, and placed about 185~\textmu m in front of the grating, with a continuous-mode microwave generator (see Fig.~\ref{fig:geom}).

\begin{figure}[htp]
\centering
\includegraphics[width=\linewidth]{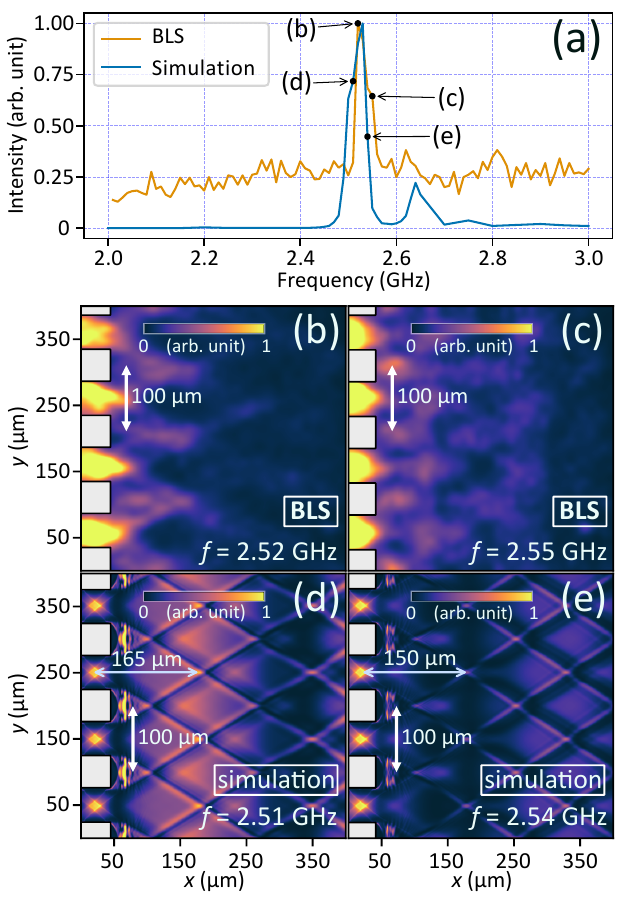}
\caption{\label{fig:caustics} 
Experimental and simulation results for a diffraction grating composed of square antidots [Fig.~\ref{fig:geom}(a)] at a magnetic field $B_\text{ext}=36$~mT. Panel (a) shows the amplitude-frequency characteristic of the excited SWs in the interference region. Panels (b) and (c) show the dynamic magnetization pattern obtained from BLS measurements, while panels (d) and (e) display the SW intensity distribution [see Eq.~(\ref{eq:energy})] calculated by MuMax3 for $f=2.51$~GHz and $f=2.54$~GHz, respectively. The arrows indicate the distances to the nearest caustic beam intersections.}
\end{figure}

\begin{figure*}[htp]
\centering
\includegraphics[width=\linewidth]{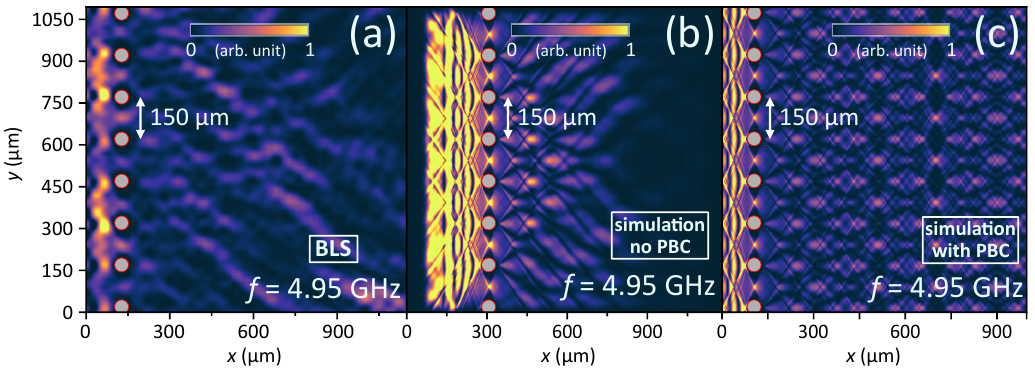}
\caption{\label{fig:self_imaging}
The SW field intensity distributions at $B_\text{ext}=98$~mT and $f=4.95$~GHz in the in-plane external magnetic field of 98~mT. Panel (a) shows the results of BLS measurements, and (b) the corresponding micromagnetic simulation, where a finite-length diffraction grating was used. In the SW intensity plot (c), PBC is applied along the $y$-axis, revealing a clear SW self-imaging effect. All the results were obtained for a diffraction grating composed of 10 circular antidots [Fig.~\ref{fig:geom}(b)] separated by 150~nm. Due to matching zooms from the experiment, the visualizations show only 8 nanodots.}
\end{figure*}

The interaction of SWs with the line of antidots was visualized using a Brillouin light scattering (BLS) spectrometer with a spatial resolution of 30~\textmu m. Measurements are made in the reflection configuration due to the opaque substrate used to mount the microwave antenna. A laser beam with a wavelength of 532~nm was scanned over the area around the line of antidots and the interference area behind the grating with 20~\textmu m step and the BLS intensity was recorded at each point. This technique provides a 2D colour map of the amplitude of the magnetostatic SWs scattered on the line of antidots [see Fig.~\ref{fig:caustics}(b,c) and Fig.~\ref{fig:self_imaging}(a)].

Simulations were carried out using MuMax3, a GPU-accelerated micromagnetic simulation software~\cite{vansteenkiste2014design}. The implemented simulation system ($100 \times 100 \times 4.5$~\textmu m$^3$) was discretized with $512 \times 512 \times 10$ computational cells, giving a size of about $195.3 \times 195.3 \times 450.0$~nm$^3$ each. The cell size clearly exceeds the length of the exchange interaction for YIG films due to computational limitations. However, according to the dispersion relation graphs (see Supplementary Materials, Fig.~S1), the discrepancies are insignificant for small wavevectors, justifying the use of numerical methods for this research.

To replicate the SWs excited by the microstrip in the experiment, in micromagnetic simulations we employ the dynamic magnetic field ${\bf h}(x,t)$, homogeneous in the area of width $w=50$~\textmu m along the $x$-axis (extended along the $y$ direction and across the film thickness) and placed at $x_0=185$~\textmu m before the grating. The microwave field is expressed as:
\begin{equation}
{\bf h}(x,t) = [h_0,0,h_0]\sin(2\pi f t),
\end{equation}
where $h_0$ is the amplitude of the dynamic magnetic field ($h_0 = 0.0014\;B_{\mathrm{ext}}$).

The distribution of the SW intensity within the magnetic material is determined by averaging the magnetization component $m_z$ over the thickness of the material (along the $z$-axis) and integrating its squared value over time $t$. The SW intensity $I$ is then quantified by the equation:
\begin{equation}
    I = \int \langle m_z(t) \rangle_z^2 dt.
    \label{eq:energy}
\end{equation}
This method effectively captures the spatial intensity distribution of SWs in the material, allowing for direct comparison with BLS measurement results.

In the simulations we use geometries shown in Fig.~\ref{fig:geom}, with dimensions matching those of our experiment. In all cases, we implemented absorbing boundary conditions along the $x$-axis (at the end of a system). These are characterized by an exponential increase in the damping factor $\alpha$, approaching a maximum value of 1 at the edges. Additionally, to better visualize the diffraction patterns formed behind the grating, periodic boundary conditions (PBC) along the $y$-axis were used in most simulations (indicated if not). It enabled the imitation of an infinitely wide film and an array of antidots, extending a given SW diffraction field to longer distances. The YIG is characterized by the following magnetic parameters: saturation magnetization $M_\mathrm{s}=139$~kA/m, exchange constant $A_\mathrm{ex}=4$~pJ/m, $|\gamma|=176$~GHz$\cdot$rad/T and reduced Gilbert damping $\alpha=1\times10^{-7}$. 

Previous investigations of the self-imaging of SWs, manifesting the Talbot effect~\cite{Golebiewski2020, Golebiewski_multimode}, have extensively utilized the out-of-plane magnetic field configuration, i.e., forward volume SWs. In this orientation, the isofrequency lines tend to be circular due to the symmetry of the applied field with respect to the film plane, leading to the isotropic propagation characteristics of SWs. This isotropic nature facilitates the formation of distinct interference patterns, critical for achieving the clear SW Talbot effect, also in relatively thick YIG films, considered in this paper (see Fig.~S2 in Supplementary Materials). However, the out-of-plane magnetic field configuration has several disadvantages when applied to thin magnetic films. A primary limitation is the need for high magnetic field values or strong out-of-plane anisotropy to saturate the magnetization. They impose practical limitations on the generation and control of large magnetic fields in experiments, and thus in potential applications where compact and efficient magnetic field generation is critical. These challenges underscore the importance of exploring alternative configurations, such as the in-plane magnetic field orientation.

\begin{figure}[htp]
\centering
\includegraphics[width=\linewidth]{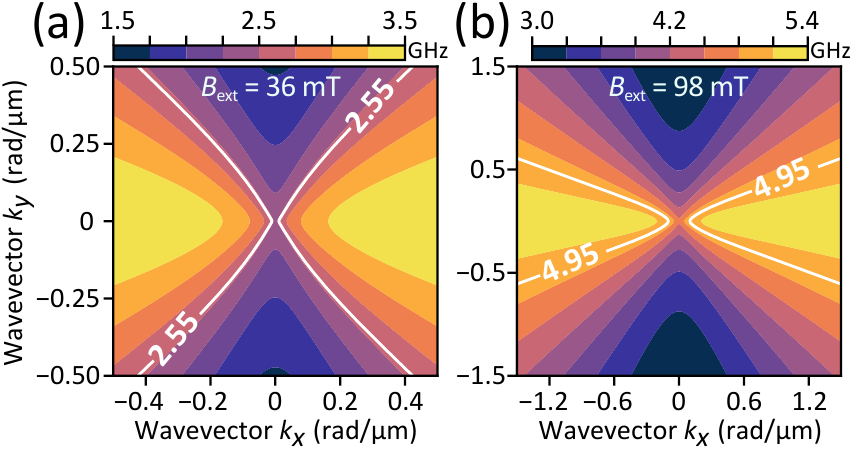}
\caption{\label{fig:iso_freq}
Diagram of the SW dispersion relation for an external magnetic field of (a) $B_\text{ext}=36$~mT and (b) $B_\text{ext}=98$~mT applied in-plane of the YIG film and along the $y$ direction. The highlighted isofrequency lines correspond to the values used in the study and are obtained from that analytical model of Ref.~\cite{Kalinikos1986} with free boundary conditions.}
\end{figure}

Avoiding the high magnetic field requirements associated with the forward volume SW configuration is a key feature of the DE configuration. However, this naturally leads to anisotropic SW propagation due to the asymmetry introduced by the in-plane magnetic field~\cite{Kalinikos1986}. This anisotropy results in hyperbolic isofrequency contours at small wavenumbers (Fig.~\ref{fig:iso_freq}), offering directional control over SW propagation~\cite{Ogasawara2023}. When studying the dynamics of SWs in ferromagnetic films with a thickness of 4.5~\textmu m in an in-plane magnetic field configuration, the interplay between the magnetic field magnitude, SWs frequency, and diffraction grating period together is of primary importance on the interference pattern, that determines the appearance of caustic waves or the Talbot effect.

At lower magnetic field values and SW frequencies, the system with hyperbolic isofrequency lines predominantly forms caustic beams post-interaction with a narrow slot\cite{Veerakumar2006,Schneider2010}. When a plane SW passes through a diffraction grating, its wavefront is modulated, resulting in a discrete spectrum of wavevectors\cite{Mansfeld2012}. As these diffracted waves propagate through the magnetic medium, their paths are influenced by the strong anisotropic dispersion relation (Fig.~\ref{fig:iso_freq}). It causes the SWs to focus along certain trajectories, leading to the convergence of the waves at specific focal points or lines\cite{Gieniusz2013,Muralidhar2021}. This convergence is the fundamental mechanism behind the formation of caustic beams, where the wave intensity is significantly amplified.

A theoretical model for caustic beam formation is usually based on analysis of the function $f(\mathbf{k})$ and the angle $\phi$, representing the orientation of the group velocity $\mathbf{v}_\text{gr}$ relative to the external magnetic field $\mathbf{B}_\text{ext}$ \cite{Buttner2000,Schneider2010, Wartelle2023}:
\begin{equation}
    \phi=\arctan(v_{\text{gr},y}/v_{\text{gr},x})=-\arctan(dk_x/dk_y).
    \label{eq:phi}
\end{equation}
At each point, the SW group velocity is indicated by the normal to the isofrequency curve. Caustic rays are formed when the direction of the group velocity, defined by the angle $\phi$, remains constant for SWs with different wavevectors $\mathbf{k}$. This specific condition for caustic beam emergence is mathematically expressed as $\frac{d\phi}{dk_x}=0$, which ensures that SWs with different wavevectors maintain a uniform group velocity direction, resulting in wave convergence. When a plane wave passes through a periodic structure, such as the one-dimensional antidot array in our study, it acquires a distinct transverse wavevector component $k_y$ quantized as multiples of $2\pi/d$. Depending on the change in angle $\phi$ for the surface wavefront and its intensity for a discrete set of $k_y$, one of the aforementioned patterns -- caustic rays, diffractive self-imaging effect, or a combination of them -- is obtained.

Fig.~\ref{fig:caustics}(a) shows the amplitude-frequency characteristic of the sample with square antidots at $B_\text{ext}=36$~mT. In the experiment, which is well reproduced in the simulations, frequencies in the range of $2.5$~GHz to $2.6$~GHz exhibit the most efficient excitation. In addition, these frequencies are slightly higher than the ferromagnetic resonance frequencies, which can be attributed to the selectivity of the antenna in exciting SWs with non-zero wavevectors. 

\begin{figure}[htp]
\centering
\includegraphics[width=\linewidth]{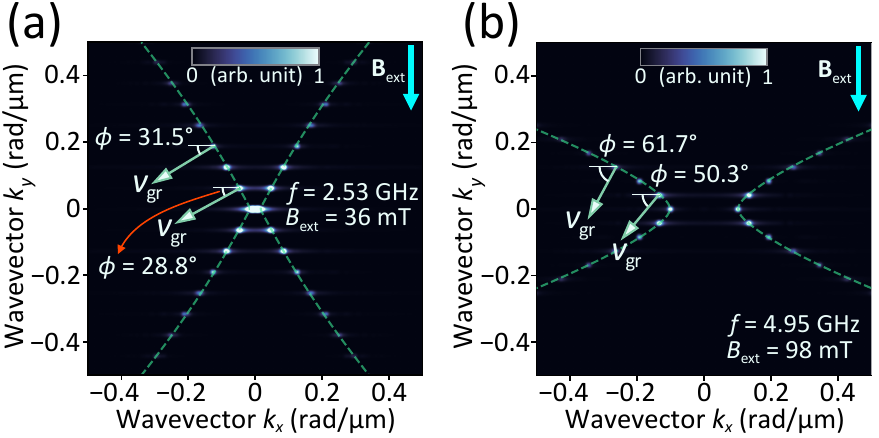}
\caption{\label{fig:isofreq_fft}
Reciprocal space maps of the simulated SW intensity distribution after passing the grating for the two cases analyzed: (a) the 100~\textmu m period square grating and (b) the 150~\textmu m period circular grating. The graphs illustrate the group velocity vectors for selected peaks and their slope $\phi$ (see Eq.~\ref{eq:phi}) with respect to the SW propagation direction. Interpolated iso-frequency contours are shown as green dashed lines.}
\end{figure}

In Fig.~\ref{fig:caustics}, we see the results of both the experimental studies (b,c) and micromagnetic simulations (d,e) of plane wave propagation through a diffraction grating with square antidots [scheme shown in Fig.~\ref{fig:geom}(a)]. The choice of frequencies in the simulations slightly shifted from those of the experiment (by 10~MHz each) is determined by the best fit of the rhombic pattern to that of the experiment. In addition, the goal was also to show the difference in the distance at which the caustic beams intersect for two different, excited SW frequencies from the region of high excitation [marked with arrows in Fig.~\ref{fig:caustics}(d,e)]. For example, in Fig.~\ref{fig:caustics}(a) we see that at 2.55~GHz the experimental intensity is high, but in the simulations, it is already quenched, hence the need for a slight offset.
The chosen SW frequencies ($f=2.51$~GHz and $f=2.54$~GHz) fit for the high transmission range at $B_\text{ext}=36$~mT [Fig.~\ref{fig:caustics}(a)], and allow us to observe the caustic, non-diffractive propagation of the beam after passing through the obstacles.

As seen in Fig.~\ref{fig:iso_freq}(a), the isofrequency lines are nearly straight. It is confirmed by a two-dimensional Fourier transform of the SW signal performed over the simulation area behind the grating shown in Fig.~\ref{fig:isofreq_fft}(a). It allows to extract of the individual sets of wavevectors involved in the formation of the diffraction images~\cite{Wartelle2023}. Each of the visible intensity peaks corresponds to a packet of wavevectors with very similar angles and group velocity vector values that can independently produce a caustic effect. In the experiment and simulations, Fig.~\ref{fig:caustics}, we see a blend of several such packets with varying intensities, producing a complex combination of caustic and interference effects. In the spectra from Fig.~\ref{fig:isofreq_fft}(a) at 2.53~GHz, we see that the difference between the angles $\Delta\phi$ for the group velocity vectors $\mathbf{v}_\text{gr}$ for the second and fourth diffraction spots (range of most intensive high order diffraction) is only 2.7$^\circ$, which explains the clear observation of the caustic effect for this configuration. The discrepancy between them is closely related to the difference in the diffraction angle of the passed waves, and its value is a function of the excited plane wave's frequency (see also Fig.~S3 in the Supplementary Material).

It is expected that at higher frequencies, as the caustic conditions weaken, the system will begin to exhibit behavior similar to the Talbot effect. This will be the case if the group velocity direction angles differ for each excited wavevector, i.e., the isofrequency curve becomes more parabolic than hyperbolic, as shown in Fig.~\ref{fig:iso_freq}(b) and also in Fig.~\ref{fig:isofreq_fft}(b) for $B_\text{ext}=98$~mT at $f=4.95$~GHz ($\lambda=84.4$~\textmu m). For these parameters, the self-imaging effect shall be more pronounced due to larger differences in the group velocity angles than in the pure caustic case, still being below the caustic angle \(\phi_\text{C} = 69.3^\circ \) (obtained from the linearly interpolated dispersion relation from Fig.~\ref{fig:iso_freq} for $k_x$ up to 3.5~rad/\textmu m). For \( B_{\text{ext}} = 36 \, \text{mT} \) and a frequency of \( 2.53 \, \text{GHz} \) this angle was only \( \phi_\text{C} = 44.7^\circ \). This combination of factors allows us to observe periodic diffraction patterns of SWs, and the Talbot-like effect. Experimentally, however, these observations were only possible for the circular antidote sample with 150~\textmu m period. Therefore, in the following analysis, we present the experimental and simulation results for this separation at \( B_{\text{ext}} = 98 \, \text{mT} \) and \( 4.95 \, \text{GHz} \).

As shown in Fig.~\ref{fig:self_imaging}, initially, after passing through the diffraction grating, the SW beams still have a caustic nature. However, as the distance from the grating increases, these beams gradually form the interference pattern. This is due to the increasing influence of larger wavevectors with distance from the source, which no longer satisfies the causticity condition described above [see Fig.~\ref{fig:isofreq_fft}(b)], where $\Delta \phi=11.4^\circ$ between the spots of the second and fourth diffraction order. Nevertheless, it is clear, especially from the micromagnetic simulations (see Fig.~S4 in the Supplementary Material) that caustic beams and self-imaging coexist even at large distances from the grating. The Talbot-like effect observed in the simulations [Fig.~\ref{fig:self_imaging}(b,c)] agrees well with the BLS measurements shown in Fig.~\ref{fig:self_imaging}(a). Experimentally, the interference pattern is limited to a triangular shape due to the finite number (10) of circular antidots used in the sample. This limitation is replicated in simulations [Fig.~\ref{fig:self_imaging}(b)] with 10 antidots and absorbing boundaries at the edges.

Comparing the interference patterns obtained in micromagnetic simulations for diffraction on circular and square antidots for the same period, field, and frequency, we found that the self-imaging patterns are very similar (see Fig.~S5 in the Supplementary Material). However, a more intense signal is obtained for the circular antidots. This difference between the shapes can be attributed to the difference in the demagnetization fields near the antidots\cite{Grafe2020}. Furthermore, this may be an additional reason for the difficulties in measuring self-imaging patterns in the sample discussed earlier for higher frequencies.

In summary, this paper provides a numerical and experimental analysis of SW self-imaging by a diffraction grating in the in-plane magnetized thin YIG films, traversing the intricate transition from caustic beam formation to the Talbot-like effect, and demonstrating the nuanced interplay between anisotropic SW dynamics, magnetic field configurations, and diffraction grating geometries. We explore SW behavior at static magnetic field values of 36~mT and 98~mT and, sequentially, excited SW frequencies of (2.51~GHz) 2.53~GHz and 4.95~GHz, respectively.

At lower fields and frequencies, we observed the formation of caustic beams, a phenomenon emerging due to the strong anisotropic nature of SW propagation in the DE configuration. As we increase the magnetic field strength and SW frequency, a transition to diffractive self-imaging patterns is observed. This switch from caustic beams to the interference self-imaging, not only exemplifies complex wave dynamics in magnonic systems but also enhances our understanding of SW diffraction mechanisms in ferromagnetic films.

\begin{acknowledgments}
This work was supported by the National Science Centre Poland project OPUS-LAP no 2020/39/I/ST3/02413 and project M-ERA.NET 3 no 2022/04/Y/ST5/00164. The authors thank Andrzej Maziewski for valuable discussions.
\end{acknowledgments}

\section*{Author declarations}

\subsection*{Conflict of Interest}
The authors have no conflicts to disclose.

\subsection*{Author Contributions}
\textbf{Uladzislau Makartsou:} Conceptualization (equal); Data Curation (lead); Formal Analysis (equal); Methodology (lead);
Software (lead); Validation (lead); Visualization (lead); Writing/Original Draft Preparation (equal). \textbf{Mateusz Gołębiewski:} Conceptualization (equal); Methodology (equal);
Supervision (equal); Validation (equal); Visualization (supporting); Writing/Original Draft Preparation (lead); Writing/Review \& Editing (equal). \textbf{Urszula Guzowska:}
Data Curation (equal); Validation (equal); Visualization (equal); Writing/Review \& Editing (supporting). \textbf{Alexander Stognij:} Data Curation (equal); Resources (equal); \textbf{Ryszard Gieniusz:} Conceptualization (equal); Data Curation (equal); Writing/Review \& Editing (supporting). \textbf{Maciej Krawczyk:} Conceptualization (equal); Formal Analysis (equal); Funding Acquisition (lead); Methodology (equal); Supervision (lead); Writing/Original Draft Preparation (equal); Writing/Review \& Editing (equal).

\section*{Data Availability Statement}
The data that support the findings of this study are openly available on Zenodo at \url{https://doi.org/10.5281/zenodo.10962369}.

\bibliography{bibliography}

\providecommand{\noopsort}[1]{}\providecommand{\singleletter}[1]{#1}%
\begin{thebibliography}{46}%
\makeatletter
\providecommand \@ifxundefined [1]{%
 \@ifx{#1\undefined}
}%
\providecommand \@ifnum [1]{%
 \ifnum #1\expandafter \@firstoftwo
 \else \expandafter \@secondoftwo
 \fi
}%
\providecommand \@ifx [1]{%
 \ifx #1\expandafter \@firstoftwo
 \else \expandafter \@secondoftwo
 \fi
}%
\providecommand \natexlab [1]{#1}%
\providecommand \enquote  [1]{``#1''}%
\providecommand \bibnamefont  [1]{#1}%
\providecommand \bibfnamefont [1]{#1}%
\providecommand \citenamefont [1]{#1}%
\providecommand \href@noop [0]{\@secondoftwo}%
\providecommand \href [0]{\begingroup \@sanitize@url \@href}%
\providecommand \@href[1]{\@@startlink{#1}\@@href}%
\providecommand \@@href[1]{\endgroup#1\@@endlink}%
\providecommand \@sanitize@url [0]{\catcode `\\12\catcode `\$12\catcode `\&12\catcode `\#12\catcode `\^12\catcode `\_12\catcode `\%12\relax}%
\providecommand \@@startlink[1]{}%
\providecommand \@@endlink[0]{}%
\providecommand \url  [0]{\begingroup\@sanitize@url \@url }%
\providecommand \@url [1]{\endgroup\@href {#1}{\urlprefix }}%
\providecommand \urlprefix  [0]{URL }%
\providecommand \Eprint [0]{\href }%
\providecommand \doibase [0]{https://doi.org/}%
\providecommand \selectlanguage [0]{\@gobble}%
\providecommand \bibinfo  [0]{\@secondoftwo}%
\providecommand \bibfield  [0]{\@secondoftwo}%
\providecommand \translation [1]{[#1]}%
\providecommand \BibitemOpen [0]{}%
\providecommand \bibitemStop [0]{}%
\providecommand \bibitemNoStop [0]{.\EOS\space}%
\providecommand \EOS [0]{\spacefactor3000\relax}%
\providecommand \BibitemShut  [1]{\csname bibitem#1\endcsname}%
\let\auto@bib@innerbib\@empty
\bibitem [{\citenamefont {Stancil}\ and\ \citenamefont {Prabhakar}(2009)}]{Stancil2009}%
  \BibitemOpen
  \bibfield  {author} {\bibinfo {author} {\bibfnamefont {D.~D.}\ \bibnamefont {Stancil}}\ and\ \bibinfo {author} {\bibfnamefont {A.}~\bibnamefont {Prabhakar}},\ }\href {https://doi.org/10.1007/978-0-387-77865-5} {\emph {\bibinfo {title} {Spin waves: Theory and applications}}}\ (\bibinfo  {publisher} {Springer US},\ \bibinfo {year} {2009})\ pp.\ \bibinfo {pages} {1--355}\BibitemShut {NoStop}%
\bibitem [{\citenamefont {Gurevich}\ and\ \citenamefont {Melkov}(1996)}]{gurevich1996book}%
  \BibitemOpen
  \bibfield  {author} {\bibinfo {author} {\bibfnamefont {A.~G.}\ \bibnamefont {Gurevich}}\ and\ \bibinfo {author} {\bibfnamefont {G.~A.}\ \bibnamefont {Melkov}},\ }\href@noop {} {\emph {\bibinfo {title} {Magnetization oscillations and waves}}}\ (\bibinfo  {publisher} {CRC Press, Boca Raton},\ \bibinfo {year} {1996})\BibitemShut {NoStop}%
\bibitem [{\citenamefont {Garcia-Sanchez}\ \emph {et~al.}(2015)\citenamefont {Garcia-Sanchez}, \citenamefont {Borys}, \citenamefont {Soucaille}, \citenamefont {Adam}, \citenamefont {Stamps},\ and\ \citenamefont {Kim}}]{Garcia-Sanchez2015NarrowWalls}%
  \BibitemOpen
  \bibfield  {author} {\bibinfo {author} {\bibfnamefont {F.}~\bibnamefont {Garcia-Sanchez}}, \bibinfo {author} {\bibfnamefont {P.}~\bibnamefont {Borys}}, \bibinfo {author} {\bibfnamefont {R.}~\bibnamefont {Soucaille}}, \bibinfo {author} {\bibfnamefont {J.~P.}\ \bibnamefont {Adam}}, \bibinfo {author} {\bibfnamefont {R.~L.}\ \bibnamefont {Stamps}},\ and\ \bibinfo {author} {\bibfnamefont {J.~V.}\ \bibnamefont {Kim}},\ }\bibfield  {title} {\enquote {\bibinfo {title} {{Narrow Magnonic Waveguides Based on Domain Walls}},}\ }\href {https://doi.org/10.1103/PhysRevLett.114.247206} {\bibfield  {journal} {\bibinfo  {journal} {Physical Review Letters}\ }\textbf {\bibinfo {volume} {114}},\ \bibinfo {pages} {247206} (\bibinfo {year} {2015})}\BibitemShut {NoStop}%
\bibitem [{\citenamefont {Wagner}\ \emph {et~al.}(2016)\citenamefont {Wagner}, \citenamefont {K{\'{a}}kay}, \citenamefont {Schultheiss}, \citenamefont {Henschke}, \citenamefont {Sebastian},\ and\ \citenamefont {Schultheiss}}]{Wagner2016MagneticNanochannels}%
  \BibitemOpen
  \bibfield  {author} {\bibinfo {author} {\bibfnamefont {K.}~\bibnamefont {Wagner}}, \bibinfo {author} {\bibfnamefont {A.}~\bibnamefont {K{\'{a}}kay}}, \bibinfo {author} {\bibfnamefont {K.}~\bibnamefont {Schultheiss}}, \bibinfo {author} {\bibfnamefont {A.}~\bibnamefont {Henschke}}, \bibinfo {author} {\bibfnamefont {T.}~\bibnamefont {Sebastian}},\ and\ \bibinfo {author} {\bibfnamefont {H.}~\bibnamefont {Schultheiss}},\ }\bibfield  {title} {\enquote {\bibinfo {title} {{Magnetic domain walls as reconfigurable spin-wave nanochannels}},}\ }\href {https://doi.org/10.1038/nnano.2015.339} {\bibfield  {journal} {\bibinfo  {journal} {Nature Nanotechnology}\ }\textbf {\bibinfo {volume} {11}},\ \bibinfo {pages} {432--436} (\bibinfo {year} {2016})}\BibitemShut {NoStop}%
\bibitem [{\citenamefont {Duerr}\ \emph {et~al.}(2012)\citenamefont {Duerr}, \citenamefont {Thurner}, \citenamefont {Topp}, \citenamefont {Huber},\ and\ \citenamefont {Grundler}}]{Duerr2012EnhancedWaveguide}%
  \BibitemOpen
  \bibfield  {author} {\bibinfo {author} {\bibfnamefont {G.}~\bibnamefont {Duerr}}, \bibinfo {author} {\bibfnamefont {K.}~\bibnamefont {Thurner}}, \bibinfo {author} {\bibfnamefont {J.}~\bibnamefont {Topp}}, \bibinfo {author} {\bibfnamefont {R.}~\bibnamefont {Huber}},\ and\ \bibinfo {author} {\bibfnamefont {D.}~\bibnamefont {Grundler}},\ }\bibfield  {title} {\enquote {\bibinfo {title} {{Enhanced transmission through squeezed modes in a self-cladding magnonic waveguide}},}\ }\href {https://doi.org/10.1103/PhysRevLett.108.227202} {\bibfield  {journal} {\bibinfo  {journal} {Physical Review Letters}\ }\textbf {\bibinfo {volume} {108}},\ \bibinfo {pages} {227202} (\bibinfo {year} {2012})}\BibitemShut {NoStop}%
\bibitem [{\citenamefont {Lan}\ \emph {et~al.}(2015)\citenamefont {Lan}, \citenamefont {Yu}, \citenamefont {Wu},\ and\ \citenamefont {Xiao}}]{Lan2015Spin-WaveDiode}%
  \BibitemOpen
  \bibfield  {author} {\bibinfo {author} {\bibfnamefont {J.}~\bibnamefont {Lan}}, \bibinfo {author} {\bibfnamefont {W.}~\bibnamefont {Yu}}, \bibinfo {author} {\bibfnamefont {R.}~\bibnamefont {Wu}},\ and\ \bibinfo {author} {\bibfnamefont {J.}~\bibnamefont {Xiao}},\ }\bibfield  {title} {\enquote {\bibinfo {title} {{Spin-Wave Diode}},}\ }\href {https://doi.org/10.1103/PHYSREVX.5.041049/FIGURES/5/MEDIUM} {\bibfield  {journal} {\bibinfo  {journal} {Physical Review X}\ }\textbf {\bibinfo {volume} {5}},\ \bibinfo {pages} {041049} (\bibinfo {year} {2015})}\BibitemShut {NoStop}%
\bibitem [{\citenamefont {Annenkov}, \citenamefont {Gerus},\ and\ \citenamefont {Lock}(2018)}]{Lock2018}%
  \BibitemOpen
  \bibfield  {author} {\bibinfo {author} {\bibfnamefont {A.~Y.}\ \bibnamefont {Annenkov}}, \bibinfo {author} {\bibfnamefont {S.~V.}\ \bibnamefont {Gerus}},\ and\ \bibinfo {author} {\bibfnamefont {E.~H.}\ \bibnamefont {Lock}},\ }\bibfield  {title} {\enquote {\bibinfo {title} {Superdirectional beam of surface spin wave},}\ }\href {https://doi.org/10.1209/0295-5075/123/44003} {\bibfield  {journal} {\bibinfo  {journal} {Europhysics Letters}\ }\textbf {\bibinfo {volume} {123}},\ \bibinfo {pages} {44003} (\bibinfo {year} {2018})}\BibitemShut {NoStop}%
\bibitem [{\citenamefont {Pirro}\ \emph {et~al.}(2021)\citenamefont {Pirro}, \citenamefont {Vasyuchka}, \citenamefont {Serga},\ and\ \citenamefont {Hillebrands}}]{Pirro2021AdvancesMagnonics}%
  \BibitemOpen
  \bibfield  {author} {\bibinfo {author} {\bibfnamefont {P.}~\bibnamefont {Pirro}}, \bibinfo {author} {\bibfnamefont {V.~I.}\ \bibnamefont {Vasyuchka}}, \bibinfo {author} {\bibfnamefont {A.~A.}\ \bibnamefont {Serga}},\ and\ \bibinfo {author} {\bibfnamefont {B.}~\bibnamefont {Hillebrands}},\ }\bibfield  {title} {\enquote {\bibinfo {title} {{Advances in coherent magnonics}},}\ }\href {https://doi.org/10.1038/s41578-021-00332-w} {\bibfield  {journal} {\bibinfo  {journal} {Nature Reviews Materials}\ }\textbf {\bibinfo {volume} {6}},\ \bibinfo {pages} {1114--1135} (\bibinfo {year} {2021})}\BibitemShut {NoStop}%
\bibitem [{\citenamefont {Whitehead}\ \emph {et~al.}(2019)\citenamefont {Whitehead}, \citenamefont {Horsley}, \citenamefont {Philbin},\ and\ \citenamefont {Kruglyak}}]{graded_index_lenses}%
  \BibitemOpen
  \bibfield  {author} {\bibinfo {author} {\bibfnamefont {N.~J.}\ \bibnamefont {Whitehead}}, \bibinfo {author} {\bibfnamefont {S.~A.~R.}\ \bibnamefont {Horsley}}, \bibinfo {author} {\bibfnamefont {T.~G.}\ \bibnamefont {Philbin}},\ and\ \bibinfo {author} {\bibfnamefont {V.~V.}\ \bibnamefont {Kruglyak}},\ }\bibfield  {title} {\enquote {\bibinfo {title} {Graded index lenses for spin wave steering},}\ }\href {https://doi.org/10.1103/PhysRevB.100.094404} {\bibfield  {journal} {\bibinfo  {journal} {Phys. Rev. B}\ }\textbf {\bibinfo {volume} {100}},\ \bibinfo {pages} {094404} (\bibinfo {year} {2019})}\BibitemShut {NoStop}%
\bibitem [{\citenamefont {Kiechle}\ \emph {et~al.}(2023)\citenamefont {Kiechle}, \citenamefont {Papp}, \citenamefont {Mendisch}, \citenamefont {Ahrens}, \citenamefont {Golibrzuch}, \citenamefont {Bernstein}, \citenamefont {Porod}, \citenamefont {Csaba},\ and\ \citenamefont {Becherer}}]{Kiechle2023}%
  \BibitemOpen
  \bibfield  {author} {\bibinfo {author} {\bibfnamefont {M.}~\bibnamefont {Kiechle}}, \bibinfo {author} {\bibfnamefont {A.}~\bibnamefont {Papp}}, \bibinfo {author} {\bibfnamefont {S.}~\bibnamefont {Mendisch}}, \bibinfo {author} {\bibfnamefont {V.}~\bibnamefont {Ahrens}}, \bibinfo {author} {\bibfnamefont {M.}~\bibnamefont {Golibrzuch}}, \bibinfo {author} {\bibfnamefont {G.~H.}\ \bibnamefont {Bernstein}}, \bibinfo {author} {\bibfnamefont {W.}~\bibnamefont {Porod}}, \bibinfo {author} {\bibfnamefont {G.}~\bibnamefont {Csaba}},\ and\ \bibinfo {author} {\bibfnamefont {M.}~\bibnamefont {Becherer}},\ }\bibfield  {title} {\enquote {\bibinfo {title} {Spin-wave optics in {YIG} realized by ion-beam irradiation},}\ }\href {https://doi.org/https://doi.org/10.1002/smll.202207293} {\bibfield  {journal} {\bibinfo  {journal} {Small}\ }\textbf {\bibinfo {volume} {19}},\ \bibinfo {pages} {2207293} (\bibinfo {year} {2023})}\BibitemShut {NoStop}%
\bibitem [{\citenamefont {Whitehead}\ \emph {et~al.}(2018)\citenamefont {Whitehead}, \citenamefont {Horsley}, \citenamefont {Philbin},\ and\ \citenamefont {Kruglyak}}]{Luneberg}%
  \BibitemOpen
  \bibfield  {author} {\bibinfo {author} {\bibfnamefont {N.~J.}\ \bibnamefont {Whitehead}}, \bibinfo {author} {\bibfnamefont {S.~A.~R.}\ \bibnamefont {Horsley}}, \bibinfo {author} {\bibfnamefont {T.~G.}\ \bibnamefont {Philbin}},\ and\ \bibinfo {author} {\bibfnamefont {V.~V.}\ \bibnamefont {Kruglyak}},\ }\bibfield  {title} {\enquote {\bibinfo {title} {A luneburg lens for spin waves},}\ }\href {https://doi.org/10.1063/1.5049470} {\bibfield  {journal} {\bibinfo  {journal} {Applied Physics Letters}\ }\textbf {\bibinfo {volume} {113}},\ \bibinfo {pages} {212404} (\bibinfo {year} {2018})}\BibitemShut {NoStop}%
\bibitem [{\citenamefont {Vogel}, \citenamefont {Hillebrands},\ and\ \citenamefont {von Freymann}(2020)}]{SW_Fourier}%
  \BibitemOpen
  \bibfield  {author} {\bibinfo {author} {\bibfnamefont {M.}~\bibnamefont {Vogel}}, \bibinfo {author} {\bibfnamefont {B.}~\bibnamefont {Hillebrands}},\ and\ \bibinfo {author} {\bibfnamefont {G.}~\bibnamefont {von Freymann}},\ }\bibfield  {title} {\enquote {\bibinfo {title} {Optical elements for anisotropic spin-wave propagation},}\ }\href {https://doi.org/10.1063/5.0018519} {\bibfield  {journal} {\bibinfo  {journal} {Applied Physics Letters}\ }\textbf {\bibinfo {volume} {116}},\ \bibinfo {pages} {262404} (\bibinfo {year} {2020})}\BibitemShut {NoStop}%
\bibitem [{\citenamefont {Demidov}\ \emph {et~al.}(2008)\citenamefont {Demidov}, \citenamefont {Demokritov}, \citenamefont {Rott}, \citenamefont {Krzysteczko},\ and\ \citenamefont {Reiss}}]{Dem08}%
  \BibitemOpen
  \bibfield  {author} {\bibinfo {author} {\bibfnamefont {V.~E.}\ \bibnamefont {Demidov}}, \bibinfo {author} {\bibfnamefont {S.~O.}\ \bibnamefont {Demokritov}}, \bibinfo {author} {\bibfnamefont {K.}~\bibnamefont {Rott}}, \bibinfo {author} {\bibfnamefont {P.}~\bibnamefont {Krzysteczko}},\ and\ \bibinfo {author} {\bibfnamefont {G.}~\bibnamefont {Reiss}},\ }\bibfield  {title} {\enquote {\bibinfo {title} {Mode interference and periodic self-focusing of spin waves in permalloy microstripes},}\ }\href {https://doi.org/10.1103/PhysRevB.77.064406} {\bibfield  {journal} {\bibinfo  {journal} {Phys. Rev. B}\ }\textbf {\bibinfo {volume} {77}},\ \bibinfo {pages} {064406} (\bibinfo {year} {2008})}\BibitemShut {NoStop}%
\bibitem [{\citenamefont {Mansfeld}\ \emph {et~al.}(2012{\natexlab{a}})\citenamefont {Mansfeld}, \citenamefont {Topp}, \citenamefont {Martens}, \citenamefont {Toedt}, \citenamefont {Hansen}, \citenamefont {Heitmann},\ and\ \citenamefont {Mendach}}]{Man12}%
  \BibitemOpen
  \bibfield  {author} {\bibinfo {author} {\bibfnamefont {S.}~\bibnamefont {Mansfeld}}, \bibinfo {author} {\bibfnamefont {J.}~\bibnamefont {Topp}}, \bibinfo {author} {\bibfnamefont {K.}~\bibnamefont {Martens}}, \bibinfo {author} {\bibfnamefont {J.~N.}\ \bibnamefont {Toedt}}, \bibinfo {author} {\bibfnamefont {W.}~\bibnamefont {Hansen}}, \bibinfo {author} {\bibfnamefont {D.}~\bibnamefont {Heitmann}},\ and\ \bibinfo {author} {\bibfnamefont {S.}~\bibnamefont {Mendach}},\ }\bibfield  {title} {\enquote {\bibinfo {title} {Spin wave diffraction and perfect imaging of a grating},}\ }\href {https://doi.org/10.1103/PhysRevLett.108.047204} {\bibfield  {journal} {\bibinfo  {journal} {Phys. Rev. Lett.}\ }\textbf {\bibinfo {volume} {108}},\ \bibinfo {pages} {047204} (\bibinfo {year} {2012}{\natexlab{a}})}\BibitemShut {NoStop}%
\bibitem [{\citenamefont {Khomeriki}(2004)}]{Kho04}%
  \BibitemOpen
  \bibfield  {author} {\bibinfo {author} {\bibfnamefont {R.}~\bibnamefont {Khomeriki}},\ }\bibfield  {title} {\enquote {\bibinfo {title} {Self-focusing magnetostatic beams in thin magnetic films},}\ }\href@noop {} {\bibfield  {journal} {\bibinfo  {journal} {Eur. Phys. J. B - Condensed Matter and Complex Systems}\ }\textbf {\bibinfo {volume} {41}},\ \bibinfo {pages} {219--222} (\bibinfo {year} {2004})}\BibitemShut {NoStop}%
\bibitem [{\citenamefont {Gieniusz}\ \emph {et~al.}(2017)\citenamefont {Gieniusz}, \citenamefont {Gruszecki}, \citenamefont {Krawczyk}, \citenamefont {Guzowska}, \citenamefont {Stognij},\ and\ \citenamefont {Maziewski}}]{gieniusz2017switching}%
  \BibitemOpen
  \bibfield  {author} {\bibinfo {author} {\bibfnamefont {R.}~\bibnamefont {Gieniusz}}, \bibinfo {author} {\bibfnamefont {P.}~\bibnamefont {Gruszecki}}, \bibinfo {author} {\bibfnamefont {M.}~\bibnamefont {Krawczyk}}, \bibinfo {author} {\bibfnamefont {U.}~\bibnamefont {Guzowska}}, \bibinfo {author} {\bibfnamefont {A.}~\bibnamefont {Stognij}},\ and\ \bibinfo {author} {\bibfnamefont {A.}~\bibnamefont {Maziewski}},\ }\bibfield  {title} {\enquote {\bibinfo {title} {The switching of strong spin wave beams in patterned garnet films},}\ }\href@noop {} {\bibfield  {journal} {\bibinfo  {journal} {Sci. Rep.}\ }\textbf {\bibinfo {volume} {7}},\ \bibinfo {pages} {8771} (\bibinfo {year} {2017})}\BibitemShut {NoStop}%
\bibitem [{\citenamefont {K{\"o}rner}, \citenamefont {Stigloher},\ and\ \citenamefont {Back}(2017)}]{korner2017excitation}%
  \BibitemOpen
  \bibfield  {author} {\bibinfo {author} {\bibfnamefont {H.}~\bibnamefont {K{\"o}rner}}, \bibinfo {author} {\bibfnamefont {J.}~\bibnamefont {Stigloher}},\ and\ \bibinfo {author} {\bibfnamefont {C.}~\bibnamefont {Back}},\ }\bibfield  {title} {\enquote {\bibinfo {title} {Excitation and tailoring of diffractive spin-wave beams in nife using nonuniform microwave antennas},}\ }\href@noop {} {\bibfield  {journal} {\bibinfo  {journal} {Phys. Rev. B}\ }\textbf {\bibinfo {volume} {96}},\ \bibinfo {pages} {100401} (\bibinfo {year} {2017})}\BibitemShut {NoStop}%
\bibitem [{\citenamefont {Gieniusz}\ \emph {et~al.}(2014)\citenamefont {Gieniusz}, \citenamefont {Bessonov}, \citenamefont {Guzowska}, \citenamefont {Stognii},\ and\ \citenamefont {Maziewski}}]{Gieniusz2014}%
  \BibitemOpen
  \bibfield  {author} {\bibinfo {author} {\bibfnamefont {R.}~\bibnamefont {Gieniusz}}, \bibinfo {author} {\bibfnamefont {V.~D.}\ \bibnamefont {Bessonov}}, \bibinfo {author} {\bibfnamefont {U.}~\bibnamefont {Guzowska}}, \bibinfo {author} {\bibfnamefont {A.~I.}\ \bibnamefont {Stognii}},\ and\ \bibinfo {author} {\bibfnamefont {A.}~\bibnamefont {Maziewski}},\ }\bibfield  {title} {\enquote {\bibinfo {title} {{An antidot array as an edge for total non-reflection of spin waves in yttrium iron garnet films}},}\ }\href {https://doi.org/10.1063/1.4867026} {\bibfield  {journal} {\bibinfo  {journal} {Applied Physics Letters}\ }\textbf {\bibinfo {volume} {104}},\ \bibinfo {pages} {082412} (\bibinfo {year} {2014})}\BibitemShut {NoStop}%
\bibitem [{\citenamefont {{T}albot}(1836)}]{Talbot36}%
  \BibitemOpen
  \bibfield  {author} {\bibinfo {author} {\bibfnamefont {H.}~\bibnamefont {{T}albot}},\ }\bibfield  {title} {\enquote {\bibinfo {title} {Lxxvi. {F}acts relating to optical science. no. iv},}\ }\href {https://doi.org/10.1080/14786443608649032} {\bibfield  {journal} {\bibinfo  {journal} {The London, Edinburgh, and Dublin Philosophical Magazine and Journal of Science}\ }\textbf {\bibinfo {volume} {9}},\ \bibinfo {pages} {401--407} (\bibinfo {year} {1836})}\BibitemShut {NoStop}%
\bibitem [{\citenamefont {Rayleigh}(1881)}]{Rayleigh81}%
  \BibitemOpen
  \bibfield  {author} {\bibinfo {author} {\bibfnamefont {L.}~\bibnamefont {Rayleigh}},\ }\bibfield  {title} {\enquote {\bibinfo {title} {On copying diffraction gratings and on some phenomenon connected therewith},}\ }\href {https://doi.org/10.1080/14786448108626995} {\bibfield  {journal} {\bibinfo  {journal} {Phil. Mag.}\ }\textbf {\bibinfo {volume} {11}},\ \bibinfo {pages} {196} (\bibinfo {year} {1881})}\BibitemShut {NoStop}%
\bibitem [{\citenamefont {Wen}, \citenamefont {Zhang},\ and\ \citenamefont {Xiao}(2013)}]{Wen13}%
  \BibitemOpen
  \bibfield  {author} {\bibinfo {author} {\bibfnamefont {J.}~\bibnamefont {Wen}}, \bibinfo {author} {\bibfnamefont {Y.}~\bibnamefont {Zhang}},\ and\ \bibinfo {author} {\bibfnamefont {M.}~\bibnamefont {Xiao}},\ }\bibfield  {title} {\enquote {\bibinfo {title} {The {T}albot effect: recent advances in classical optics, nonlinear optics, and quantum optics},}\ }\href {https://doi.org/10.1364/AOP.5.000083} {\bibfield  {journal} {\bibinfo  {journal} {Adv. Opt. Photon.}\ }\textbf {\bibinfo {volume} {5}},\ \bibinfo {pages} {83--130} (\bibinfo {year} {2013})}\BibitemShut {NoStop}%
\bibitem [{\citenamefont {Bravin}, \citenamefont {Coan},\ and\ \citenamefont {Suortti}(2012)}]{Bravin_2012}%
  \BibitemOpen
  \bibfield  {author} {\bibinfo {author} {\bibfnamefont {A.}~\bibnamefont {Bravin}}, \bibinfo {author} {\bibfnamefont {P.}~\bibnamefont {Coan}},\ and\ \bibinfo {author} {\bibfnamefont {P.}~\bibnamefont {Suortti}},\ }\bibfield  {title} {\enquote {\bibinfo {title} {X-ray phase-contrast imaging: from pre-clinical applications towards clinics},}\ }\href {https://doi.org/10.1088/0031-9155/58/1/r1} {\bibfield  {journal} {\bibinfo  {journal} {Physics in Medicine and Biology}\ }\textbf {\bibinfo {volume} {58}},\ \bibinfo {pages} {R1--R35} (\bibinfo {year} {2012})}\BibitemShut {NoStop}%
\bibitem [{\citenamefont {Sato}(2014)}]{Sat14}%
  \BibitemOpen
  \bibfield  {author} {\bibinfo {author} {\bibfnamefont {T.}~\bibnamefont {Sato}},\ }\bibfield  {title} {\enquote {\bibinfo {title} {Focus position and depth of two-dimensional patterning by {T}albot effect lithography},}\ }\href {https://doi.org/https://doi.org/10.1016/j.mee.2014.05.019} {\bibfield  {journal} {\bibinfo  {journal} {Microelectronic Engineering}\ }\textbf {\bibinfo {volume} {123}},\ \bibinfo {pages} {80 -- 83} (\bibinfo {year} {2014})},\ \bibinfo {note} {nano Lithography 2013}\BibitemShut {NoStop}%
\bibitem [{\citenamefont {{Zhou}}\ \emph {et~al.}(2016)\citenamefont {{Zhou}}, \citenamefont {{Liu}}, \citenamefont {{Deng}}, \citenamefont {{Xie}},\ and\ \citenamefont {{Chan}}}]{Zho16}%
  \BibitemOpen
  \bibfield  {author} {\bibinfo {author} {\bibfnamefont {S.}~\bibnamefont {{Zhou}}}, \bibinfo {author} {\bibfnamefont {J.}~\bibnamefont {{Liu}}}, \bibinfo {author} {\bibfnamefont {Q.}~\bibnamefont {{Deng}}}, \bibinfo {author} {\bibfnamefont {C.}~\bibnamefont {{Xie}}},\ and\ \bibinfo {author} {\bibfnamefont {M.}~\bibnamefont {{Chan}}},\ }\bibfield  {title} {\enquote {\bibinfo {title} {Depth-of-focus determination for {T}albot lithography of large-scale free-standing periodic features},}\ }\href {https://doi.org/10.1109/LPT.2016.2601639} {\bibfield  {journal} {\bibinfo  {journal} {IEEE Photon. Technol. Lett.}\ }\textbf {\bibinfo {volume} {28}},\ \bibinfo {pages} {2491--2494} (\bibinfo {year} {2016})}\BibitemShut {NoStop}%
\bibitem [{\citenamefont {Vetter}\ \emph {et~al.}(2018)\citenamefont {Vetter}, \citenamefont {Kirner}, \citenamefont {Opalevs}, \citenamefont {Scholz}, \citenamefont {Leisching}, \citenamefont {Scharf}, \citenamefont {Noell}, \citenamefont {Rockstuhl},\ and\ \citenamefont {Voelkel}}]{Vet18}%
  \BibitemOpen
  \bibfield  {author} {\bibinfo {author} {\bibfnamefont {A.}~\bibnamefont {Vetter}}, \bibinfo {author} {\bibfnamefont {R.}~\bibnamefont {Kirner}}, \bibinfo {author} {\bibfnamefont {D.}~\bibnamefont {Opalevs}}, \bibinfo {author} {\bibfnamefont {M.}~\bibnamefont {Scholz}}, \bibinfo {author} {\bibfnamefont {P.}~\bibnamefont {Leisching}}, \bibinfo {author} {\bibfnamefont {T.}~\bibnamefont {Scharf}}, \bibinfo {author} {\bibfnamefont {W.}~\bibnamefont {Noell}}, \bibinfo {author} {\bibfnamefont {C.}~\bibnamefont {Rockstuhl}},\ and\ \bibinfo {author} {\bibfnamefont {R.}~\bibnamefont {Voelkel}},\ }\bibfield  {title} {\enquote {\bibinfo {title} {Printing sub-micron structures using {T}albot mask-aligner lithography with a 193 nm cw laser light source},}\ }\href {https://doi.org/10.1364/OE.26.022218} {\bibfield  {journal} {\bibinfo  {journal} {Opt. Express}\ }\textbf {\bibinfo {volume} {26}},\ \bibinfo {pages} {22218--22233} (\bibinfo {year} {2018})}\BibitemShut {NoStop}%
\bibitem [{\citenamefont {Bigourd}\ \emph {et~al.}(2008)\citenamefont {Bigourd}, \citenamefont {Chatel}, \citenamefont {Schleich},\ and\ \citenamefont {Girard}}]{Big08}%
  \BibitemOpen
  \bibfield  {author} {\bibinfo {author} {\bibfnamefont {D.}~\bibnamefont {Bigourd}}, \bibinfo {author} {\bibfnamefont {B.}~\bibnamefont {Chatel}}, \bibinfo {author} {\bibfnamefont {W.~P.}\ \bibnamefont {Schleich}},\ and\ \bibinfo {author} {\bibfnamefont {B.}~\bibnamefont {Girard}},\ }\bibfield  {title} {\enquote {\bibinfo {title} {Factorization of numbers with the temporal {T}albot effect: Optical implementation by a sequence of shaped ultrashort pulses},}\ }\href {https://doi.org/10.1103/PhysRevLett.100.030202} {\bibfield  {journal} {\bibinfo  {journal} {Phys. Rev. Lett.}\ }\textbf {\bibinfo {volume} {100}},\ \bibinfo {pages} {030202} (\bibinfo {year} {2008})}\BibitemShut {NoStop}%
\bibitem [{\citenamefont {Far\'{\i}as}\ \emph {et~al.}(2015)\citenamefont {Far\'{\i}as}, \citenamefont {de~Melo}, \citenamefont {Milman},\ and\ \citenamefont {Walborn}}]{Far15}%
  \BibitemOpen
  \bibfield  {author} {\bibinfo {author} {\bibfnamefont {O.~J.}\ \bibnamefont {Far\'{\i}as}}, \bibinfo {author} {\bibfnamefont {F.}~\bibnamefont {de~Melo}}, \bibinfo {author} {\bibfnamefont {P.}~\bibnamefont {Milman}},\ and\ \bibinfo {author} {\bibfnamefont {S.~P.}\ \bibnamefont {Walborn}},\ }\bibfield  {title} {\enquote {\bibinfo {title} {Quantum information processing by weaving quantum {T}albot carpets},}\ }\href {https://doi.org/10.1103/PhysRevA.91.062328} {\bibfield  {journal} {\bibinfo  {journal} {Phys. Rev. A}\ }\textbf {\bibinfo {volume} {91}},\ \bibinfo {pages} {062328} (\bibinfo {year} {2015})}\BibitemShut {NoStop}%
\bibitem [{\citenamefont {Sawada}\ and\ \citenamefont {Walborn}(2018)}]{Saw18}%
  \BibitemOpen
  \bibfield  {author} {\bibinfo {author} {\bibfnamefont {K.}~\bibnamefont {Sawada}}\ and\ \bibinfo {author} {\bibfnamefont {S.~P.}\ \bibnamefont {Walborn}},\ }\bibfield  {title} {\enquote {\bibinfo {title} {Experimental quantum information processing with the {T}albot effect},}\ }\href {https://doi.org/10.1088/2040-8986/aac5c1} {\bibfield  {journal} {\bibinfo  {journal} {J. Opt.}\ }\textbf {\bibinfo {volume} {20}},\ \bibinfo {pages} {075201} (\bibinfo {year} {2018})}\BibitemShut {NoStop}%
\bibitem [{\citenamefont {Dennis}, \citenamefont {Zheludev},\ and\ \citenamefont {de~Abajo}(2007)}]{Den07}%
  \BibitemOpen
  \bibfield  {author} {\bibinfo {author} {\bibfnamefont {M.~R.}\ \bibnamefont {Dennis}}, \bibinfo {author} {\bibfnamefont {N.~I.}\ \bibnamefont {Zheludev}},\ and\ \bibinfo {author} {\bibfnamefont {F.~J.~G.}\ \bibnamefont {de~Abajo}},\ }\bibfield  {title} {\enquote {\bibinfo {title} {The plasmon {T}albot effect},}\ }\href {https://doi.org/10.1364/OE.15.009692} {\bibfield  {journal} {\bibinfo  {journal} {Opt. Express}\ }\textbf {\bibinfo {volume} {15}},\ \bibinfo {pages} {9692--9700} (\bibinfo {year} {2007})}\BibitemShut {NoStop}%
\bibitem [{\citenamefont {Sungar}\ \emph {et~al.}(2018)\citenamefont {Sungar}, \citenamefont {Sharpe}, \citenamefont {Pilgram}, \citenamefont {Bernard},\ and\ \citenamefont {Tambasco}}]{Sun18}%
  \BibitemOpen
  \bibfield  {author} {\bibinfo {author} {\bibfnamefont {N.}~\bibnamefont {Sungar}}, \bibinfo {author} {\bibfnamefont {J.}~\bibnamefont {Sharpe}}, \bibinfo {author} {\bibfnamefont {J.}~\bibnamefont {Pilgram}}, \bibinfo {author} {\bibfnamefont {J.}~\bibnamefont {Bernard}},\ and\ \bibinfo {author} {\bibfnamefont {L.}~\bibnamefont {Tambasco}},\ }\bibfield  {title} {\enquote {\bibinfo {title} {Faraday-{T}albot effect: Alternating phase and circular arrays},}\ }\href@noop {} {\bibfield  {journal} {\bibinfo  {journal} {Chaos: An Interdisciplinary Journal of Nonlinear Science}\ }\textbf {\bibinfo {volume} {28}},\ \bibinfo {pages} {096101} (\bibinfo {year} {2018})}\BibitemShut {NoStop}%
\bibitem [{\citenamefont {Bakman}\ \emph {et~al.}(2019)\citenamefont {Bakman}, \citenamefont {Fishman}, \citenamefont {Fink}, \citenamefont {Fort},\ and\ \citenamefont {Wildeman}}]{Bakman19}%
  \BibitemOpen
  \bibfield  {author} {\bibinfo {author} {\bibfnamefont {A.}~\bibnamefont {Bakman}}, \bibinfo {author} {\bibfnamefont {S.}~\bibnamefont {Fishman}}, \bibinfo {author} {\bibfnamefont {M.}~\bibnamefont {Fink}}, \bibinfo {author} {\bibfnamefont {E.}~\bibnamefont {Fort}},\ and\ \bibinfo {author} {\bibfnamefont {S.}~\bibnamefont {Wildeman}},\ }\bibfield  {title} {\enquote {\bibinfo {title} {Observation of the {T}albot effect with water waves},}\ }\href@noop {} {\bibfield  {journal} {\bibinfo  {journal} {Am. J. Phys.}\ }\textbf {\bibinfo {volume} {87}},\ \bibinfo {pages} {38--43} (\bibinfo {year} {2019})}\BibitemShut {NoStop}%
\bibitem [{\citenamefont {Gao}\ \emph {et~al.}(2016)\citenamefont {Gao}, \citenamefont {Estrecho}, \citenamefont {Li}, \citenamefont {Egorov}, \citenamefont {Ma}, \citenamefont {Winkler}, \citenamefont {Kamp}, \citenamefont {Schneider}, \citenamefont {H\"ofling}, \citenamefont {Truscott},\ and\ \citenamefont {Ostrovskaya}}]{Gao16}%
  \BibitemOpen
  \bibfield  {author} {\bibinfo {author} {\bibfnamefont {T.}~\bibnamefont {Gao}}, \bibinfo {author} {\bibfnamefont {E.}~\bibnamefont {Estrecho}}, \bibinfo {author} {\bibfnamefont {G.}~\bibnamefont {Li}}, \bibinfo {author} {\bibfnamefont {O.~A.}\ \bibnamefont {Egorov}}, \bibinfo {author} {\bibfnamefont {X.}~\bibnamefont {Ma}}, \bibinfo {author} {\bibfnamefont {K.}~\bibnamefont {Winkler}}, \bibinfo {author} {\bibfnamefont {M.}~\bibnamefont {Kamp}}, \bibinfo {author} {\bibfnamefont {C.}~\bibnamefont {Schneider}}, \bibinfo {author} {\bibfnamefont {S.}~\bibnamefont {H\"ofling}}, \bibinfo {author} {\bibfnamefont {A.~G.}\ \bibnamefont {Truscott}},\ and\ \bibinfo {author} {\bibfnamefont {E.~A.}\ \bibnamefont {Ostrovskaya}},\ }\bibfield  {title} {\enquote {\bibinfo {title} {{T}albot effect for exciton polaritons},}\ }\href {https://doi.org/10.1103/PhysRevLett.117.097403} {\bibfield  {journal} {\bibinfo  {journal} {Phys. Rev. Lett.}\ }\textbf {\bibinfo {volume} {117}},\ \bibinfo {pages} {097403} (\bibinfo {year}
  {2016})}\BibitemShut {NoStop}%
\bibitem [{\citenamefont {Goł\c{e}biewski}\ \emph {et~al.}(2020)\citenamefont {Goł\c{e}biewski}, \citenamefont {Gruszecki}, \citenamefont {Krawczyk},\ and\ \citenamefont {Serebryannikov}}]{Golebiewski2020}%
  \BibitemOpen
  \bibfield  {author} {\bibinfo {author} {\bibfnamefont {M.}~\bibnamefont {Goł\c{e}biewski}}, \bibinfo {author} {\bibfnamefont {P.}~\bibnamefont {Gruszecki}}, \bibinfo {author} {\bibfnamefont {M.}~\bibnamefont {Krawczyk}},\ and\ \bibinfo {author} {\bibfnamefont {A.~E.}\ \bibnamefont {Serebryannikov}},\ }\bibfield  {title} {\enquote {\bibinfo {title} {Spin-wave {T}albot effect in a thin ferromagnetic film},}\ }\href {https://doi.org/10.1103/PhysRevB.102.134402} {\bibfield  {journal} {\bibinfo  {journal} {Phys. Rev. B}\ }\textbf {\bibinfo {volume} {102}},\ \bibinfo {pages} {134402} (\bibinfo {year} {2020})}\BibitemShut {NoStop}%
\bibitem [{\citenamefont {Goł\c{e}biewski}, \citenamefont {Gruszecki},\ and\ \citenamefont {Krawczyk}(2022{\natexlab{a}})}]{Golebiewski_multimode}%
  \BibitemOpen
  \bibfield  {author} {\bibinfo {author} {\bibfnamefont {M.}~\bibnamefont {Goł\c{e}biewski}}, \bibinfo {author} {\bibfnamefont {P.}~\bibnamefont {Gruszecki}},\ and\ \bibinfo {author} {\bibfnamefont {M.}~\bibnamefont {Krawczyk}},\ }\bibfield  {title} {\enquote {\bibinfo {title} {Self-imaging of spin waves in thin, multimode ferromagnetic waveguides},}\ }\href {https://doi.org/10.1109/TMAG.2022.3140280} {\bibfield  {journal} {\bibinfo  {journal} {IEEE Transactions on Magnetics}\ }\textbf {\bibinfo {volume} {58}},\ \bibinfo {pages} {1--5} (\bibinfo {year} {2022}{\natexlab{a}})}\BibitemShut {NoStop}%
\bibitem [{\citenamefont {Goł\c{e}biewski}, \citenamefont {Gruszecki},\ and\ \citenamefont {Krawczyk}(2022{\natexlab{b}})}]{Golebiewski_lookup}%
  \BibitemOpen
  \bibfield  {author} {\bibinfo {author} {\bibfnamefont {M.}~\bibnamefont {Goł\c{e}biewski}}, \bibinfo {author} {\bibfnamefont {P.}~\bibnamefont {Gruszecki}},\ and\ \bibinfo {author} {\bibfnamefont {M.}~\bibnamefont {Krawczyk}},\ }\bibfield  {title} {\enquote {\bibinfo {title} {Self-imaging based programmable spin-wave lookup tables},}\ }\href {https://doi.org/https://doi.org/10.1002/aelm.202200373} {\bibfield  {journal} {\bibinfo  {journal} {Advanced Electronic Materials}\ }\textbf {\bibinfo {volume} {8}},\ \bibinfo {pages} {2200373} (\bibinfo {year} {2022}{\natexlab{b}})}\BibitemShut {NoStop}%
\bibitem [{\citenamefont {Vansteenkiste}\ \emph {et~al.}(2014)\citenamefont {Vansteenkiste}, \citenamefont {Leliaert}, \citenamefont {Dvornik}, \citenamefont {Helsen}, \citenamefont {Garcia-Sanchez},\ and\ \citenamefont {Van~Waeyenberge}}]{vansteenkiste2014design}%
  \BibitemOpen
  \bibfield  {author} {\bibinfo {author} {\bibfnamefont {A.}~\bibnamefont {Vansteenkiste}}, \bibinfo {author} {\bibfnamefont {J.}~\bibnamefont {Leliaert}}, \bibinfo {author} {\bibfnamefont {M.}~\bibnamefont {Dvornik}}, \bibinfo {author} {\bibfnamefont {M.}~\bibnamefont {Helsen}}, \bibinfo {author} {\bibfnamefont {F.}~\bibnamefont {Garcia-Sanchez}},\ and\ \bibinfo {author} {\bibfnamefont {B.}~\bibnamefont {Van~Waeyenberge}},\ }\bibfield  {title} {\enquote {\bibinfo {title} {The design and verification of mumax3},}\ }\href@noop {} {\bibfield  {journal} {\bibinfo  {journal} {AIP Advances}\ }\textbf {\bibinfo {volume} {4}},\ \bibinfo {pages} {107133} (\bibinfo {year} {2014})}\BibitemShut {NoStop}%
\bibitem [{\citenamefont {Kalinikos}\ and\ \citenamefont {Slavin}(1986)}]{Kalinikos1986}%
  \BibitemOpen
  \bibfield  {author} {\bibinfo {author} {\bibfnamefont {B.~A.}\ \bibnamefont {Kalinikos}}\ and\ \bibinfo {author} {\bibfnamefont {A.~N.}\ \bibnamefont {Slavin}},\ }\bibfield  {title} {\enquote {\bibinfo {title} {Theory of dipole-exchange spin wave spectrum for ferromagnetic films with mixed exchange boundary conditions},}\ }\href {https://doi.org/10.1088/0022-3719/19/35/014} {\bibfield  {journal} {\bibinfo  {journal} {Journal of Physics C: Solid State Physics}\ }\textbf {\bibinfo {volume} {19}},\ \bibinfo {pages} {7013} (\bibinfo {year} {1986})}\BibitemShut {NoStop}%
\bibitem [{\citenamefont {Ogasawara}(2023)}]{Ogasawara2023}%
  \BibitemOpen
  \bibfield  {author} {\bibinfo {author} {\bibfnamefont {T.}~\bibnamefont {Ogasawara}},\ }\bibfield  {title} {\enquote {\bibinfo {title} {Time-resolved vector-field imaging of spin-wave propagation in permalloy stripes using wide-field magneto-optical {K}err microscopy},}\ }\href {https://doi.org/10.1103/PhysRevApplied.20.024010} {\bibfield  {journal} {\bibinfo  {journal} {Phys. Rev. Appl.}\ }\textbf {\bibinfo {volume} {20}},\ \bibinfo {pages} {024010} (\bibinfo {year} {2023})}\BibitemShut {NoStop}%
\bibitem [{\citenamefont {Veerakumar}\ and\ \citenamefont {Camley}(2006)}]{Veerakumar2006}%
  \BibitemOpen
  \bibfield  {author} {\bibinfo {author} {\bibfnamefont {V.}~\bibnamefont {Veerakumar}}\ and\ \bibinfo {author} {\bibfnamefont {R.~E.}\ \bibnamefont {Camley}},\ }\bibfield  {title} {\enquote {\bibinfo {title} {Magnon focusing in thin ferromagnetic films},}\ }\href {https://doi.org/10.1103/PhysRevB.74.214401} {\bibfield  {journal} {\bibinfo  {journal} {Phys. Rev. B}\ }\textbf {\bibinfo {volume} {74}},\ \bibinfo {pages} {214401} (\bibinfo {year} {2006})}\BibitemShut {NoStop}%
\bibitem [{\citenamefont {Schneider}\ \emph {et~al.}(2010)\citenamefont {Schneider}, \citenamefont {Serga}, \citenamefont {Chumak}, \citenamefont {Sandweg}, \citenamefont {Trudel}, \citenamefont {Wolff}, \citenamefont {Kostylev}, \citenamefont {Tiberkevich}, \citenamefont {Slavin},\ and\ \citenamefont {Hillebrands}}]{Schneider2010}%
  \BibitemOpen
  \bibfield  {author} {\bibinfo {author} {\bibfnamefont {T.}~\bibnamefont {Schneider}}, \bibinfo {author} {\bibfnamefont {A.~A.}\ \bibnamefont {Serga}}, \bibinfo {author} {\bibfnamefont {A.~V.}\ \bibnamefont {Chumak}}, \bibinfo {author} {\bibfnamefont {C.~W.}\ \bibnamefont {Sandweg}}, \bibinfo {author} {\bibfnamefont {S.}~\bibnamefont {Trudel}}, \bibinfo {author} {\bibfnamefont {S.}~\bibnamefont {Wolff}}, \bibinfo {author} {\bibfnamefont {M.~P.}\ \bibnamefont {Kostylev}}, \bibinfo {author} {\bibfnamefont {V.~S.}\ \bibnamefont {Tiberkevich}}, \bibinfo {author} {\bibfnamefont {A.~N.}\ \bibnamefont {Slavin}},\ and\ \bibinfo {author} {\bibfnamefont {B.}~\bibnamefont {Hillebrands}},\ }\bibfield  {title} {\enquote {\bibinfo {title} {Nondiffractive subwavelength wave beams in a medium with externally controlled anisotropy},}\ }\href {https://doi.org/10.1103/PhysRevLett.104.197203} {\bibfield  {journal} {\bibinfo  {journal} {Phys. Rev. Lett.}\ }\textbf {\bibinfo {volume} {104}},\ \bibinfo {pages} {197203} (\bibinfo
  {year} {2010})}\BibitemShut {NoStop}%
\bibitem [{\citenamefont {Mansfeld}\ \emph {et~al.}(2012{\natexlab{b}})\citenamefont {Mansfeld}, \citenamefont {Topp}, \citenamefont {Martens}, \citenamefont {Toedt}, \citenamefont {Hansen}, \citenamefont {Heitmann},\ and\ \citenamefont {Mendach}}]{Mansfeld2012}%
  \BibitemOpen
  \bibfield  {author} {\bibinfo {author} {\bibfnamefont {S.}~\bibnamefont {Mansfeld}}, \bibinfo {author} {\bibfnamefont {J.}~\bibnamefont {Topp}}, \bibinfo {author} {\bibfnamefont {K.}~\bibnamefont {Martens}}, \bibinfo {author} {\bibfnamefont {J.~N.}\ \bibnamefont {Toedt}}, \bibinfo {author} {\bibfnamefont {W.}~\bibnamefont {Hansen}}, \bibinfo {author} {\bibfnamefont {D.}~\bibnamefont {Heitmann}},\ and\ \bibinfo {author} {\bibfnamefont {S.}~\bibnamefont {Mendach}},\ }\bibfield  {title} {\enquote {\bibinfo {title} {Spin wave diffraction and perfect imaging of a grating},}\ }\href {https://doi.org/10.1103/PhysRevLett.108.047204} {\bibfield  {journal} {\bibinfo  {journal} {Phys. Rev. Lett.}\ }\textbf {\bibinfo {volume} {108}},\ \bibinfo {pages} {047204} (\bibinfo {year} {2012}{\natexlab{b}})}\BibitemShut {NoStop}%
\bibitem [{\citenamefont {Gieniusz}\ \emph {et~al.}(2013)\citenamefont {Gieniusz}, \citenamefont {Ulrichs}, \citenamefont {Bessonov}, \citenamefont {Guzowska}, \citenamefont {Stognii},\ and\ \citenamefont {Maziewski}}]{Gieniusz2013}%
  \BibitemOpen
  \bibfield  {author} {\bibinfo {author} {\bibfnamefont {R.}~\bibnamefont {Gieniusz}}, \bibinfo {author} {\bibfnamefont {H.}~\bibnamefont {Ulrichs}}, \bibinfo {author} {\bibfnamefont {V.~D.}\ \bibnamefont {Bessonov}}, \bibinfo {author} {\bibfnamefont {U.}~\bibnamefont {Guzowska}}, \bibinfo {author} {\bibfnamefont {A.~I.}\ \bibnamefont {Stognii}},\ and\ \bibinfo {author} {\bibfnamefont {A.}~\bibnamefont {Maziewski}},\ }\bibfield  {title} {\enquote {\bibinfo {title} {{Single antidot as a passive way to create caustic spin-wave beams in yttrium iron garnet films}},}\ }\href {https://doi.org/10.1063/1.4795293} {\bibfield  {journal} {\bibinfo  {journal} {Applied Physics Letters}\ }\textbf {\bibinfo {volume} {102}},\ \bibinfo {pages} {102409} (\bibinfo {year} {2013})}\BibitemShut {NoStop}%
\bibitem [{\citenamefont {Muralidhar}\ \emph {et~al.}(2021)\citenamefont {Muralidhar}, \citenamefont {Khymyn}, \citenamefont {Awad}, \citenamefont {Alem\'an}, \citenamefont {Hanstorp},\ and\ \citenamefont {\AA{}kerman}}]{Muralidhar2021}%
  \BibitemOpen
  \bibfield  {author} {\bibinfo {author} {\bibfnamefont {S.}~\bibnamefont {Muralidhar}}, \bibinfo {author} {\bibfnamefont {R.}~\bibnamefont {Khymyn}}, \bibinfo {author} {\bibfnamefont {A.~A.}\ \bibnamefont {Awad}}, \bibinfo {author} {\bibfnamefont {A.}~\bibnamefont {Alem\'an}}, \bibinfo {author} {\bibfnamefont {D.}~\bibnamefont {Hanstorp}},\ and\ \bibinfo {author} {\bibfnamefont {J.}~\bibnamefont {\AA{}kerman}},\ }\bibfield  {title} {\enquote {\bibinfo {title} {Femtosecond laser pulse driven caustic spin wave beams},}\ }\href {https://doi.org/10.1103/PhysRevLett.126.037204} {\bibfield  {journal} {\bibinfo  {journal} {Phys. Rev. Lett.}\ }\textbf {\bibinfo {volume} {126}},\ \bibinfo {pages} {037204} (\bibinfo {year} {2021})}\BibitemShut {NoStop}%
\bibitem [{\citenamefont {B{\"u}ttner}\ \emph {et~al.}(2000)\citenamefont {B{\"u}ttner}, \citenamefont {Bauer}, \citenamefont {Demokritov}, \citenamefont {Hillebrands}, \citenamefont {Kivshar}, \citenamefont {Grimalsky}, \citenamefont {Rapoport},\ and\ \citenamefont {Slavin}}]{Buttner2000}%
  \BibitemOpen
  \bibfield  {author} {\bibinfo {author} {\bibfnamefont {O.}~\bibnamefont {B{\"u}ttner}}, \bibinfo {author} {\bibfnamefont {M.}~\bibnamefont {Bauer}}, \bibinfo {author} {\bibfnamefont {S.~O.}\ \bibnamefont {Demokritov}}, \bibinfo {author} {\bibfnamefont {B.}~\bibnamefont {Hillebrands}}, \bibinfo {author} {\bibfnamefont {Y.~S.}\ \bibnamefont {Kivshar}}, \bibinfo {author} {\bibfnamefont {V.}~\bibnamefont {Grimalsky}}, \bibinfo {author} {\bibfnamefont {Y.}~\bibnamefont {Rapoport}},\ and\ \bibinfo {author} {\bibfnamefont {A.~N.}\ \bibnamefont {Slavin}},\ }\bibfield  {title} {\enquote {\bibinfo {title} {Linear and nonlinear diffraction of dipolar spin waves in yttrium iron garnet films observed by space- and time-resolved {B}rillouin light scattering},}\ }\href {https://doi.org/10.1103/PhysRevB.61.11576} {\bibfield  {journal} {\bibinfo  {journal} {Phys. Rev. B}\ }\textbf {\bibinfo {volume} {61}},\ \bibinfo {pages} {11576--11587} (\bibinfo {year} {2000})}\BibitemShut {NoStop}%
\bibitem [{\citenamefont {Wartelle}\ \emph {et~al.}(2023)\citenamefont {Wartelle}, \citenamefont {Vilsmeier}, \citenamefont {Taniguchi},\ and\ \citenamefont {Back}}]{Wartelle2023}%
  \BibitemOpen
  \bibfield  {author} {\bibinfo {author} {\bibfnamefont {A.}~\bibnamefont {Wartelle}}, \bibinfo {author} {\bibfnamefont {F.}~\bibnamefont {Vilsmeier}}, \bibinfo {author} {\bibfnamefont {T.}~\bibnamefont {Taniguchi}},\ and\ \bibinfo {author} {\bibfnamefont {C.~H.}\ \bibnamefont {Back}},\ }\bibfield  {title} {\enquote {\bibinfo {title} {Caustic spin wave beams in soft thin films: Properties and classification},}\ }\href {https://doi.org/10.1103/PhysRevB.107.144431} {\bibfield  {journal} {\bibinfo  {journal} {Phys. Rev. B}\ }\textbf {\bibinfo {volume} {107}},\ \bibinfo {pages} {144431} (\bibinfo {year} {2023})}\BibitemShut {NoStop}%
\bibitem [{\citenamefont {Gr{\"a}fe}\ \emph {et~al.}(2020)\citenamefont {Gr{\"a}fe}, \citenamefont {Gruszecki}, \citenamefont {Zelent}, \citenamefont {Decker}, \citenamefont {Keskinbora}, \citenamefont {Noske}, \citenamefont {Gawronski}, \citenamefont {Stoll}, \citenamefont {Weigand}, \citenamefont {Krawczyk}, \citenamefont {Back}, \citenamefont {Goering},\ and\ \citenamefont {Sch{\"u}tz}}]{Grafe2020}%
  \BibitemOpen
  \bibfield  {author} {\bibinfo {author} {\bibfnamefont {J.}~\bibnamefont {Gr{\"a}fe}}, \bibinfo {author} {\bibfnamefont {P.}~\bibnamefont {Gruszecki}}, \bibinfo {author} {\bibfnamefont {M.}~\bibnamefont {Zelent}}, \bibinfo {author} {\bibfnamefont {M.}~\bibnamefont {Decker}}, \bibinfo {author} {\bibfnamefont {K.}~\bibnamefont {Keskinbora}}, \bibinfo {author} {\bibfnamefont {M.}~\bibnamefont {Noske}}, \bibinfo {author} {\bibfnamefont {P.}~\bibnamefont {Gawronski}}, \bibinfo {author} {\bibfnamefont {H.}~\bibnamefont {Stoll}}, \bibinfo {author} {\bibfnamefont {M.}~\bibnamefont {Weigand}}, \bibinfo {author} {\bibfnamefont {M.}~\bibnamefont {Krawczyk}}, \bibinfo {author} {\bibfnamefont {C.~H.}\ \bibnamefont {Back}}, \bibinfo {author} {\bibfnamefont {E.~J.}\ \bibnamefont {Goering}},\ and\ \bibinfo {author} {\bibfnamefont {G.}~\bibnamefont {Sch{\"u}tz}},\ }\bibfield  {title} {\enquote {\bibinfo {title} {Direct observation of spin-wave focusing by a fresnel lens},}\ }\href
  {https://doi.org/10.1103/PhysRevB.102.024420} {\bibfield  {journal} {\bibinfo  {journal} {Phys. Rev. B}\ }\textbf {\bibinfo {volume} {102}},\ \bibinfo {pages} {024420} (\bibinfo {year} {2020})}\BibitemShut {NoStop}%
\end{thebibliography}%

\newpage
\clearpage
\onecolumngrid
\largeSection{SUPPLEMENTARY MATERIAL}
The supplementary material provides additional data, including numerical results of the dispersion relations and their comparison with theoretical predictions. These data confirm the accuracy of the numerical approaches and discretization methods used. In addition, it includes extended results from micromagnetic simulations of diffraction fields, specifically illustrating the Talbot effect over various magnetic field configurations, diffraction grating constants, and nanodot geometries.
\\
\\

\suppnumbering
\section{Dispersion relation and validation of the discretization in micromagnetic simulations}

\begin{figure}[hpt!]
    \centering
    \includegraphics[width=0.8\textwidth]{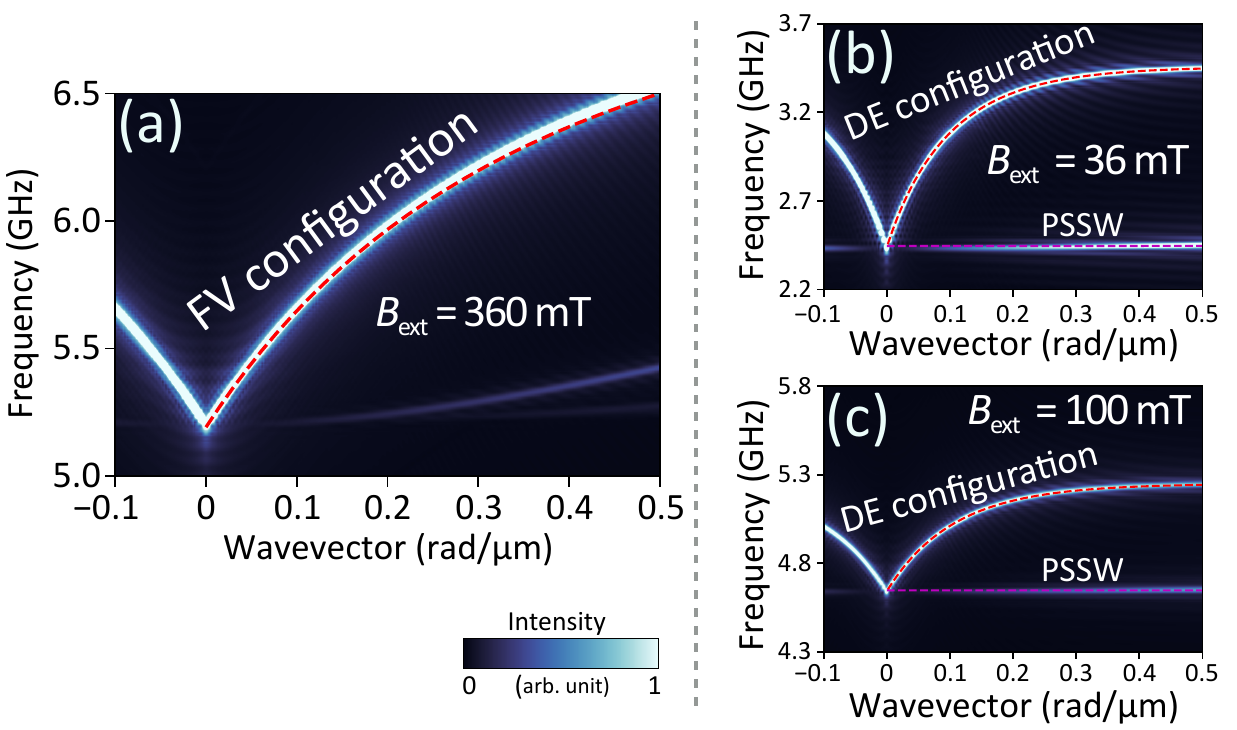}
    \caption{The dispersion relations for SWs in two different magnetic configurations, and different external magnetic field values. (a) Dispersion curve for the FV configuration with a magnetic field of 360~mT. (b) The dispersion in the DE configuration under an external field of 36~mT and PPSW signal lines are visible. (c) DE and PPSW signals for an increased field of 100~mT, revealing a notable shift in the frequency values. In all the panels, dashed lines represent the analytical calculations based on Ref.~[1].}
    \label{fig:dispersions}
\end{figure}

In Fig.~\ref{fig:dispersions} we show the dispersion relations of spin waves (SWs) for two different configurations and different values of the magnetic field obtained in numerical simulations (MuMax3) and analytical model (based on Ref.~[1]) for 4.5~\textmu m thick YIG film. Fig.~\ref{fig:dispersions}(a) shows the dispersion for the forward volume (FV) configuration (magnetic field and the magnetization perpendicular to the film plane), where discrepancies between the methods are minimal and mainly due to the coarse discretization along the film thickness. The dispersions in Figs.~\ref{fig:dispersions}(b,c) are for the Damon-Eshbach (DE) configuration (propagation direction perpendicular to the in-plane magnetization and the external magnetic field) and two different values of the external magnetic field applied. The plots also show nondispersive signals from thickness quantized SW modes, the so-called perpendicular standing SWs (PPSW). Again, we obtained a very good fit to the theoretical model for small wavevectors, which are close to zero, for validating the numerical approach used in the micromagnetic simulations.

The process of calculating the SW dispersions using Mumax\textsuperscript{3} encompassed three main steps:
\begin{enumerate}
    \item \textbf{System Initialization:} Prompting the system to relax into a stable ground state for selected configurations—either DE or FV and external magnetic field.
     \item \textbf{SW excitation:} The SWs were excited from the central part of the waveguide by applying the time- and space-dependent dynamic magnetic field $\textbf{h}$, defined as:
    $\mathbf{h}(x,t) = [h_0, h_0, h_0] \cdot \text{sinc}(2\pi k_{\text{cut}} x) \cdot \text{sinc}(2\pi f_{\text{cut}} t)$, where $h_0 = 0.015 H_{\text{ext}}$. By using this equation, we can apply broadband SW excitation within the frequency range $f \in [0, f_{\text{cut}}]$ and wavevectors along the $x$-axis $k_x \in [-k_{\text{cut}}, k_{\text{cut}}]$.
    \item \textbf{Data Analysis:} The dispersion relations were extracted by applying a 2D Fast Fourier Transformation to the space- and time-resolved magnetization data using Python with NumPy package. This analysis was further refined by evaluating the SW width profiles via the $m_x$ component across the width of the waveguides using a single frequency excitation.
    
\end{enumerate}

The dispersion relations from micromagnetic simulations and analytical calculations match precisely for the fundamental mode, as indicated in Fig.~\ref{fig:dispersions}. In the simulations with DE configuration, a low-frequency branch emerges due to quantization effects within the significant thickness (4.5~\textmu m) of the YIG sample, visible in Fig.~\ref{fig:dispersions}(b,c). The theoretical curve was calculated using the formula (22) from Ref.~[2]. The results demonstrate the agreement between MuMax3 simulations and the analytical dispersion relation based on Ref.~[1]. The simulation cell size used is $195.3 \times 195.3 \times 450.0$~nm\textsuperscript{3}. 

\section{Talbot effect in a forward volume configuration -- simulation results}

\begin{figure}[h]
\centering
\includegraphics[width=0.75\textwidth]{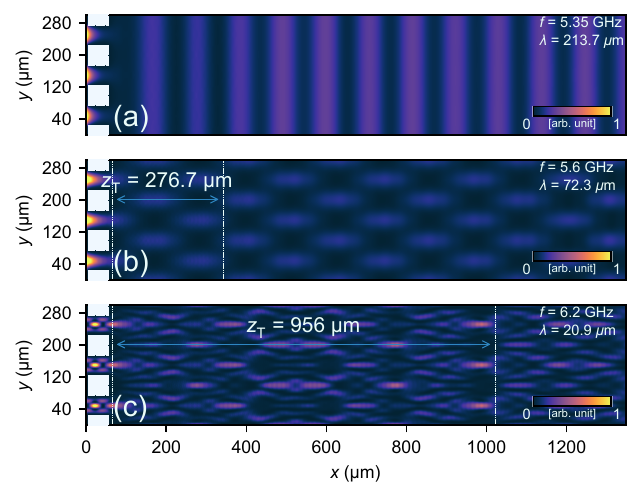}
\caption{\label{fig:FVC} The evolution of SW interference pattern after passing through a diffraction grating with a square antidote structure and a 100~\textmu m period. These micromagnetic simulations were made in an FV configuration at different frequencies (5.35, 5.6 and 6.2~GHz), and for \Bext = 100~mT. Panel (a) shows SWs with a wavelength of 213~\textmu m propagating unaffected by the obstacle, and (b) visualizes the generation of the Talbot effect after the passage of 72.3~\textmu m-long SWs through a diffraction grating. Panel (c) demonstrates the Talbot effect at a much shorter wavelength of 20.9~\textmu m, resulting in a more pronounced self-imaging effect. Periodic boundary conditions along the diffraction grating ($y$-axis) were applied to each of these simulations.}
\end{figure}

Talbot effect in magnonics has been numerically demonstrated for the configuration with the isotropic dispersion relation, i.e., in the FV configuration for the thin ferromagnetic films [see Refs. 3,4]. In Fig.~\ref{fig:FVC} we show the numerical results indicating the possibility of observing the Talbot effect in 4.5~\textmu m thick YIG film, a system considered in this paper, with a line of antidots of square shape ($50 \times 50$~\textmu m) and the period of 100~\textmu m. We can see that the diffraction of SWs with wavelengths similar to the grating period forms self-imaging patterns~(Fig.~\ref{fig:FVC}(a, b)), which is the Talbot effect and it is more clear for shorter SWs as shown in Fig.~\ref{fig:FVC}(c). In Fig.~\ref{fig:FVC}(c) we observe clear self-repeating patterns over a longer distance (the Talbot length $z_\text{T}=956$~\textmu m) than in Fig.~\ref{fig:FVC}(b) ($z_\text{T}=276.7$~\textmu m) with the SW wavelengths 20.9~\textmu m and 72.3~\textmu m, respectively.

\section{Damon-Eshbach configuration}

\begin{figure}[h]
\centering
\includegraphics[width=0.75\textwidth]{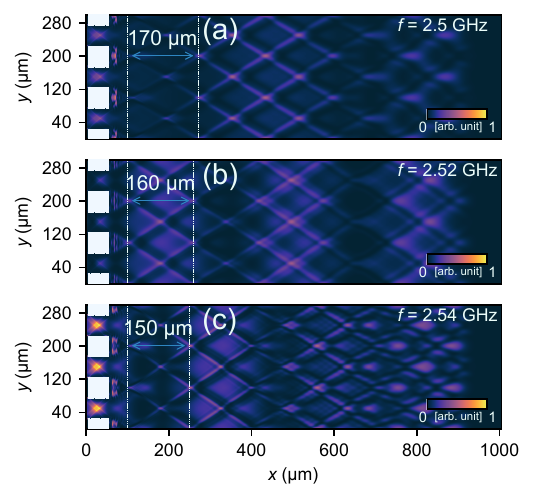}
\caption{\label{fig:DEC1} The SW interference pattern after passing through a diffraction grating with a square antidot structure and the 100~\textmu m period. The micromagnetic simulations were made in a DE configuration for frequencies 2.50, 2.52 and 2.54~GHz, and for \Bext = 36~mT.}
\end{figure}

\begin{figure}
\centering
\includegraphics[width=\textwidth]{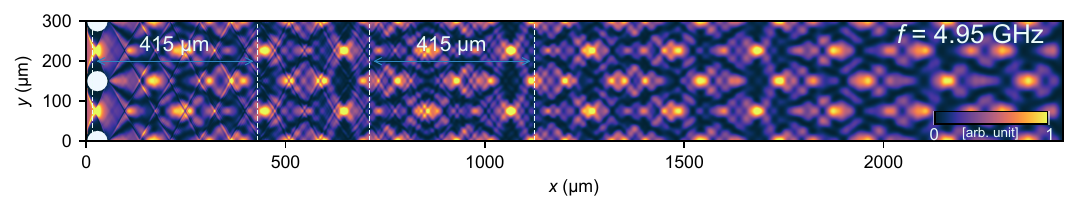}
\caption{\label{fig:DEC2} The SW interference pattern after passing through a diffraction grating with a circular antidot structure and a 150~\textmu m period. These micromagnetic simulations were made in a DE configuration for frequency 4.95~GHz, and for \Bext = 98~mT.}
\end{figure}

We show in Fig.~\ref{fig:DEC1} the extended progression of SWs following diffraction on square antidots in the in-plane magnetized YIG film in DE configuration. The analysis is conducted over an enlarged spatial domain as compared to the figures in the main text, allowing for a comprehensive observation of the SW pattern evolution.
In Fig.~\ref{fig:DEC1} the micromagnetic simulation results for SWs at 2.50, 2.52, and 2.54~GHz for \Bext = 36~mT are shown. The predominant behavior of the SWs is characterized by the caustic effect, as shown in Fig.~\ref{fig:DEC1}(a). However, as the frequency increases (wavelength shortens), the angle of caustic beam propagation increases, resulting in a decrease in the distance of crossing beams from neighboring slits along the propagation axis, from 170 to 150~\textmu m with the frequency change by 40~MHz, leading also to a more complex picture emerging, as shown in Fig.~\ref{fig:DEC1}(b,c). 
Furthermore, in Fig.~\ref{fig:DEC2} for circular antidots and 150~\textmu m period, we demonstrate the diffraction of SWs at a higher frequency (4.95~GHz) and a greater magnitude of the external magnetic field (\Bext = 98~mT). This corresponds to exciting SWs with different group velocity angles, resulting in even more intricate and periodically repeating patterns.

\section{Influence of an antidot shape on self-image}
\begin{figure}[!h]
\centering
\includegraphics[width=0.75\textwidth]{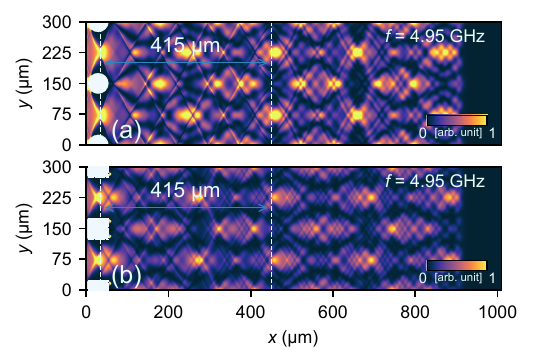}
\caption{\label{fig:DEC3}Panel (a) depicts a cut of the simulation from Fig.~\ref{fig:DEC2} featuring a circular lattice with a period of 150~\textmu m. Panel (b) illustrates the interference pattern formed by SWs passing through square antidots grating and the same period of 150~\textmu m.}
\end{figure}

We changed the shape of the antidots to investigate their effect on the SW propagation in a YIG film and the formation of the self-image pattern. We use the same parameters as in the simulation in Fig.~\ref{fig:DEC2} (lattice period 150~\textmu m, frequency 4.95~GHz, and \Bext = 98~mT), and the obtained pattern with the circular antidots as a reference image. By changing the antidot shapes from circles [Fig.~\ref{fig:DEC3}(a)] to squares [Fig.~\ref{fig:DEC3}(b)], we preserve the self-image pattern in diffraction with the same distance between repeating images, i.e., 415~\textmu m. However, the circular antidot grating provides better signal quality, resulting in the formation of clearer patterns. This difference between shapes of antidots can be attributed to the difference in the demagnetizing field near the holes Ref. [5].
\regularnumbering

\section*{References}

{\footnotesize 
\noindent [S1] A. Kalinikos and A. N. Slavin, ``Theory of dipole-exchange spin wave spectrum for ferromagnetic films with mixed exchange boundary conditions,'' \textit{Journal of Physics C: Solid State Physics} \textbf{19}, 7013 (1986).

\noindent [S2] K. Szulc, J. Kharlan, P. Bondarenko, E. V. Tartakovskaya, and M. Krawczyk, ``Impact of surface anisotropy on the spin-wave dynamics in a thin ferromagnetic film,'' \textit{Phys. Rev. B} \textbf{109}, 054430 (2024).

\noindent [S3] M. Golebiewski, P. Gruszecki, M. Krawczyk, and A. E. Serebryannikov, ``Spin-wave Talbot effect in a thin ferromagnetic film,'' \textit{Phys. Rev. B} \textbf{102}, 134402 (2020).

\noindent [S4] M. Golebiewski, P. Gruszecki, and M. Krawczyk, ``Self-imaging of spin waves in thin, multimode ferromagnetic waveguides,'' \textit{IEEE Transactions on Magnetics} \textbf{58}, 1--5 (2022).

\noindent [S5] J. Gräfe, P. Gruszecki, M. Zelent, M. Decker, K. Keskinbora, M. Noske, P. Gawronski, H. Stoll, M. Weigand, M. Krawczyk, C. H. Back, E. J. Goering, and G. Schütz, ``Direct observation of spin-wave focusing by a fresnel lens,'' \textit{Phys. Rev. B} \textbf{102}, 024420 (2020).

} 
\end{document}